\documentclass[twocolumn,aps,showpacs,showkeys,prx,superscriptaddress,reprint]{revtex4-2}
\usepackage{natbib}
\usepackage{bm}
\usepackage{bbm}
\usepackage{textcase}
\usepackage{color}
\usepackage{graphicx}
\usepackage{amsmath}
\usepackage{amssymb}
\usepackage{url}
\usepackage[colorlinks=true,citecolor=blue,hyperindex,breaklinks]{hyperref} 
\usepackage{algorithm}
\usepackage{algpseudocode}
\usepackage[font=small,justification=Justified]{caption}
\usepackage{subcaption}
\usepackage[braket, qm]{qcircuit}

\hypersetup{pdftitle={Parallel Quantum Chemistry on Noisy Intermediate-Scale Quantum Computers}}
\hypersetup{pdfauthor={Robert Schade, Carsten Bauer, Konstantin Tamoev, Lukas Mazur, Christian Plessl, Thomas D. K\"uhne}}
\hypersetup{pdfdisplaydoctitle}

\newcommand{\mat}[1]{{\boldsymbol #1}}
\newcommand{\mods}[1]{#1}

\begin{document}

\title{Parallel Quantum Chemistry on Noisy Intermediate-Scale Quantum Computers}

\author{Robert Schade}
\affiliation{Paderborn Center for Parallel Computing, Paderborn University}

\author{Carsten Bauer}
\affiliation{Paderborn Center for Parallel Computing, Paderborn University}

\author{Konstantin Tamoev}
\affiliation{Dynamics of Condensed Matter and Institute for Sustainable Systems Design, Chair of Theoretical Chemistry, Paderborn University}

\author{Lukas Mazur}
\affiliation{Paderborn Center for Parallel Computing, Paderborn University}

\author{Christian Plessl}
\affiliation{Paderborn Center for Parallel Computing, Paderborn University}
\affiliation{Department for Computer Science, Paderborn University}

\author{Thomas D. Kühne}
\affiliation{Dynamics of Condensed Matter and Institute for Sustainable Systems Design, Chair of Theoretical Chemistry, Paderborn University}

\date{\today}

\begin{abstract}
A novel parallel hybrid quantum-classical algorithm for the solution of the quantum-chemical ground-state energy problem on gate-based quantum computers is presented. 
This approach is based on the reduced density-matrix functional theory (RDMFT) formulation of the electronic structure problem. For that purpose, the density-matrix functional of the full system is decomposed into an indirectly coupled sum of density-matrix functionals for all its subsystems using the adaptive cluster approximation to RDMFT. The approximations involved in the decomposition and the adaptive cluster approximation itself can be systematically converged to the exact result.
The solutions for the density-matrix functionals of the effective subsystems involves a constrained minimization over many-particle states that are approximated by parametrized trial states on the quantum computer similarly to the variational quantum eigensolver.
The independence of the density-matrix functionals of the effective subsystems introduces a new level of parallelization and allows for the computational treatment of much larger molecules on a quantum computer with a given qubit count. In addition, for the proposed algorithm techniques are presented to reduce the qubit count, the number of quantum programs, as well as its depth. The \mods{evaluation of a density-matrix functional as the essential part of our} new approach is demonstrated for Hubbard-like systems on IBM quantum computers based on superconducting transmon qubits.
\end{abstract}


\maketitle

\section{\label{sec:introduction}Introduction}
Quantum computers have recently emerged as a powerful resource for solving computational problems that have eluded an efficient treatment on classical hardware due to their computational complexity.
A particularly promising application for quantum computing (QC) is quantum chemistry, whose focus is on solving the electronic structure problem represented by different interacting many-fermion Hamiltonians. Here, the utilization of traditional numerical methods is significantly hampered by its computational cost that scales exponentially or, for quantum Monte Carlo methods, by the infamous fermion sign problem~\cite{PhysRevLett.94.170201,PhysRevE.90.053304}. 
Consequently, quantum chemistry simulations regularly reach and push the limits of the most powerful high-performance computers.

Recently, significant research efforts have been aimed at developing novel quantum algorithms that allows one to solve electronic structure problems more efficiently. Most notably, quantum phase-estimation methods~\cite{Loyd_1996, AspuruGuzik_2005} and variational quantum eigensolvers (VQE)~\cite{Peruzzo2014, McClean_2016, kandala2017hardware,2021arXiv211105176T} for computing the (ground-state) energy of atoms and molecules have been put forward. Complementing these algorithmic advances, schemes to effectively realize quantum many-particle states on quantum hardware have been suggested~\cite{Kandala2017, OMalley2016, Shen2017}.

In this work, we assess the potential of QC for \textit{ab-initio} quantum chemistry, in which the exponential complexity lies in the quantum mechanical treatment of electrons in the calculation of the total energies and nuclear forces of many-fermion systems. Within the framework of the reduced density-matrix functional theory (RDMFT)~\cite{gilbert75_prb12_2111, Coleman_1963, Levy01121979, lieb68_prl20_1445}
we propose a hybrid quantum-classical algorithm for computing the reduced density-matrix functional (RDMF) in which the quantum mechanical expectation values of the reduced density matrix are evaluated on a quantum computer. This makes the approach very similar to the VQE in the sense that a parametrized trial state for the many-particle state is prepared on the quantum computer and
the parameters of this state are modified with a minimization algorithm running on a classical computer till the measured total energy is minimal. While the approach proposed here introduces additional equality constraints on the minimum compared to the VQE, the formulation allows for novel approximations like the adaptive cluster approximation (ACA)~\cite{PhysRevB.97.245131}.

The proposed combination of an RDMFT-based algorithm and the ACA in RDMFT approximately decomposes the RDMF of the full system into a sum of RDMFs of smaller effective systems. This on the one hand drastically reduces the required qubit count, but also makes the problem inherently parallelizable. Thus, much larger molecules can be treated on a quantum computer with a given qubit count compared to a traditional VQE. However, due to its VQE-like nature, the proposed algorithm inherits the noise tolerance and suitability for near-term quantum computers. 

The structure of the paper is as follows. \mods{First, we introduce the employed RDMFT-based approach in} Sec.~\ref{sec:RDMFT}. Afterwards, in Sec.~\ref{sec:RDMFT_QC}, we introduce our hybrid quantum-classical algorithm for computing the RDMF on quantum hardware. The subsequent sections are then concerned with the efficient implementation of this algorithm on noisy intermediate-scale quantum (NISQ) devices. Concretely, we discuss \mods{various possibilities for reducing the number of qubit operations (Sec.~\ref{sec:qcdepth}) and the number of quantum programs (Sec.~\ref{sec:programcount}).} Lastly, we showcase exemplary results for the evaluation of the RDMF of the half-filled Hubbard chain as obtained in noise-free quantum simulation,s as well as genuine simulations on NISQ hardware by IBM in Sec.~\ref{sec:results}.


\section{\label{sec:RDMFT}Reduced Density-Matrix Functional Theory}
\subsection{\label{sec:RDMFT_theo}Theoretical Framework}
RDMFT for fermionic systems uses the one-particle reduced density matrix, i.e.
\begin{equation}
    \rho^{(1)}_{\alpha,\beta}=\langle \hat c^\dagger_\beta \hat c_\alpha \rangle,
\end{equation}
with the fermionic annihilation operator $\hat c_\alpha$ for electrons in the one-particle basis state with index $\alpha$ and creation operator $\hat c^\dagger_\beta$ for electrons in the one-particle basis state with index $\beta$, as the basic quantity. Assuming a finite-dimensional one-particle basis of $N_\chi$ states, i.e. $\alpha,\beta \in \{1,...,N_\chi\}$, the one-particle reduced density matrix $\mat \rho^{(1)}\in \mathbb{C}^{N_\chi \times N_\chi}$ is hermitian.
As such, RDMFT can be seen as an extension to density functional theory (DFT) in the sense that beyond the electron density, which corresponds to the diagonal elements of $\rho^{(1)}_{\alpha,\beta}$, all elements of the one-particle reduced density matrix are considered. It is suitable for the description of strong local electronic correlations in solids and molecules because the one-particle reduced density matrix explicitly contains information about the orbital occupancies that are essential for the description of electronic correlations. Similar to DFT, RDMFT is an exact theory if the underlying functionals are not approximated.

Given a many-particle Hamiltonian of an $N$-particle system at zero temperature
\begin{equation}
    \hat H=\hat h+\hat W,
\end{equation}
which is composed of a non-interacting (single-particle) operator $\hat h=\sum_{\alpha,\beta} h_{\alpha,\beta} \hat c^\dagger_\alpha \hat c_\beta$ and a two-particle interaction operator $\hat W=\frac{1}{2} \sum_{\alpha,\beta,\gamma,\delta} U_{\alpha,\beta,\delta,\gamma} \hat c^\dagger_\alpha \hat c^\dagger_\beta \hat c_\gamma \hat c_\delta$, 
the ground-state energy $E_N(\hat h+\hat W)$ of the system can be expressed as~\cite{PhysRevB.12.2111,levy79_pnas76_6062,PhysRevB.12.2111}
\begin{equation}
\label{eq:EN}
    E_N(\hat h+\hat W)=\min_{\mat \rho^{(1)},\mat 0\leq \mat \rho^{(1)} \leq \mat 1, \mathrm{Tr}(\mat \rho^{(1)})=N } \mathrm{Tr}(\mat \rho^{(1)} \mat h)+F^{\hat W}[ \mat \rho^{(1)}].
\end{equation}
Here, the condition $\mat 0\leq \mat \rho^{(1)} \leq \mat 1$ is shorthand for the ensemble-representability of the one-particle reduced density matrix that is, its eigenvalues $f_i$ (occupations) must fulfill $0\leq f_i\leq 1$. This way, the exponential many-particle complexity of the fermionic problem is absorbed into the RDMF $F^{\hat W}[ \mat \rho^{(1)}]$, which is the analogon of the exchange-correlation functional in DFT. Since $F^{\hat W}[ \mat \rho^{(1)}]$ is universal, it only depends on the interaction and the one-particle reduced density matrix, but not on the one-particle Hamiltonian $\hat h$ that includes the external potential.

Several approximations of the RDMF have been developed and even simple variants suffice to describe strong electronic correlations beyond the abilities of local or semi-local exchange-correlation functionals~\cite{Pernal2016,MULLER1984446}. However, promising functionals have also shown pathological behaviours~\cite{PhysRevB.93.085141}. It is therefore worthwhile to take a step back to the definition of the RDMF~\cite{levy79_pnas76_6062,PhysRevB.84.205101}
\begin{equation}
\label{eq:F}
    F^{\hat W}[\mat \rho^{(1)}]= \min_{\{P_i,|\Psi_i\rangle\}\rightarrow \mat \rho^{(1)}, \sum_i P_i=1} \sum_i P_i \langle \Psi_i|\hat W |\Psi_i\rangle
\end{equation}
as the constrained minimum over an ensemble of fermionic many-particle wave functions $|\Psi_i\rangle$ with ensemble probabilities $P_i$. Beyond normalization, the major requirement is that the ensemble $\{P_i,|\Psi_i\rangle\}$ must correspond to the given one-particle reduced density matrix 
$\mat \rho^{(1)}$ via
\begin{equation}
    \rho^{(1)}_{\alpha,\beta}= \sum_i P_i \langle \Psi_i | \hat c^\dagger_\beta \hat c_\alpha | \Psi_i\rangle.
\end{equation}
Focusing on the case of a non-degenerate ground state and zero electron temperature, the minimization can be simplified to only a single many-particle wave function
\begin{equation}
\label{eq:F1}
    F^{\hat W}[\mat \rho^{(1)}]= \min_{|\Psi\rangle\rightarrow \mat \rho^{(1)}} \langle \Psi|\hat W |\Psi\rangle,
\end{equation}
with the constraints
\begin{equation}
    \rho^{(1)}_{\alpha,\beta}= \langle \Psi | \hat c^\dagger_\beta \hat c_\alpha | \Psi\rangle.
\end{equation}
Let us note in passing that the generalization of RDMFT to finite temperatures is relatively straightforward ~\cite{PhysRevB.12.2111,PhysRevA.92.052514}.

For an efficient minimization of the total energy in Eq.~\eqref{eq:EN}, the derivatives of the RDMF $F^{\hat W}[\mat \rho^{(1)}]$ with respect to the elements of the one-particle reduced density matrix $\rho^{(1)}_{\alpha,\beta}$, are required. Within our hybrid classical-quantum algorithm - to be introduced in Sec.~\ref{sec:RDMFT_QC} - these derivatives are readily available as Lagrange multipliers during the constrained minimization procedure.

\mods{Thus far, no approximations have been introduced and the constrained minimization in Eq.~\eqref{eq:F} still entails the exponential complexity of the many-particle problem. However, as we will show in the following section, the minimization problem required for the RDMF is well suited for quantum computers and can be solved via a hybrid quantum-classical algorithm on NISQ quantum hardware.}

\subsection{RDMFT-based \textit{ab-inito} Molecular Dynamics Simulations} \label{sec:RDMFT_MD}
\mods{The overall goal of this method is to perform \textit{ab-initio} molecular dynamics simulations. Hence, analytical nuclear forces are required that can be evaluated within RDMFT as} 
\begin{subequations}
\begin{align}
    \mat F_i&=-\frac{d E_N}{d \mat R_i} \label{eq:forces}\\
    &=-\frac{d }{d \mat R_i} \left(\mathrm{Tr}(\mat \rho^{(1)}_\mathrm{min}\mat h)+F^{\hat W}[\mat \rho^{(1)}_\mathrm{min}]\right),
\end{align}
\end{subequations}
with the position $\mat R_i$ of atom $i$ and the one-particle reduced density matrix $\mat \rho^{(1)}_\mathrm{min}$ in the minimum of Eq.~\eqref{eq:EN}.
The derivatives of the RDMF can be expressed as
\begin{widetext}
\begin{align}
\frac{d }{d \mat R_i}F^{\hat W}[\mat \rho^{(1)}_\mathrm{min}]&=\sum_{\alpha,\beta} \frac{\partial F^{\hat W}[\mat \rho^{(1)}_\mathrm{min}]}{\partial \rho^{(1)}_{\mathrm{min},\beta,\alpha}} \frac{\partial \rho^{(1)}_{\mathrm{min},\beta,\alpha}}{\partial \mat R_i}+
\frac{1}{2}\sum_{\alpha,\beta,\gamma,\delta} \frac{\partial U_{\alpha,\beta,\delta,\gamma}}{\partial \mat R_i} \rho^{(2)}_{\mathrm{min},\gamma\delta\alpha\beta},
\end{align}
\end{widetext}
where $\mat \rho^{(2)}_{\mathrm{min}}$ is the two-particle reduced density matrix in the minimum. The required derivatives of the RDMF with respect to the one-particle reduced density matrix are given by the Lagrange multipliers of the density-matrix constraints, i.e.
\begin{equation}
    \frac{\partial F^{\hat W}[\mat \rho^{(1)}_\mathrm{min}]}{\partial \rho^{(1)}_{\mathrm{min},\beta,\alpha}}=-\lambda_{\alpha,\beta}.
\end{equation}
Thus, an algorithm for the solution of the constrained minimization problem given in Eq.~\eqref{eq:F1} is required that produces the Lagrange multipliers. In Section~\ref{sec:RDMFT_QC} we propose to use the augmented Lagrangian approach~\cite{powellmethod,Hestenes1969} for this purpose.

\subsection{\label{sec:local}Local Approximation of the RDMF}
While for the VQE one of the only known methods to reduce the number of required qubits is to employ symmetries~\cite{bravyi2017tapering}, our RDMFT-based formulation allows additional avenues to drastically reduce the qubit count.
\mods{Starting from the general case of a RDMF $F^{\hat W}[{\mat \rho^{(1)}}]$, with a given interaction Hamiltonian $\hat W$, it is important to recognize that many of the important electronic correlation effects stem from strong local electronic interactions~\cite{gutzwiller63_prl10_159,hubbard63_prsla276_238,kanamori63_progtheorphys30_275}. We follow here the local-approximation approach of Blöchl and coworkers~\cite{PhysRevB.84.205101,Schade2017}.}

\begin{figure}[t]
    \centering
    \includegraphics[width=0.45\textwidth]{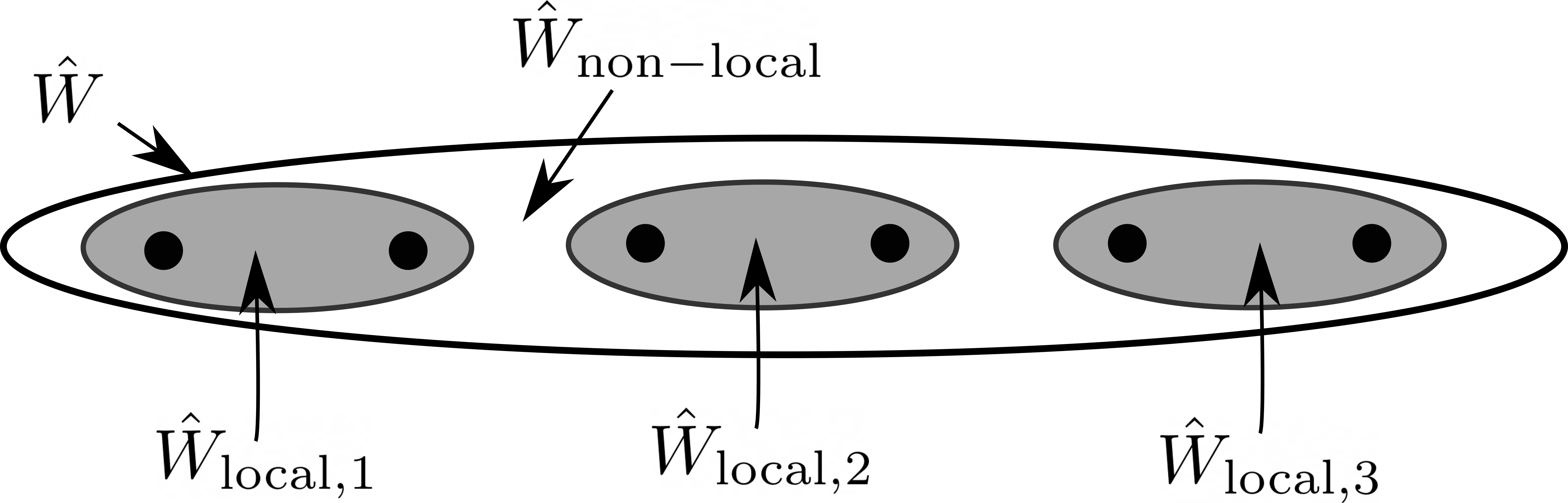}
    \caption{Schematic representation of the decomposition of the interaction Hamiltonian $\hat W$ into local terms $\hat W_{\mathrm{local},i}$ and a non-local term $\hat W_\mathrm{non-local}$ for a situation with six one-particle states and two states per local interaction term. Black dots represent the one-particle states.}
    \label{fig:local_approx}
\end{figure}

\mods{The electron-electron interaction can then be decomposed into local terms $\hat W_{\mathrm{local},i}$ and non-local terms $\hat W_\mathrm{non-local}$, i.e.,}
\begin{equation}\label{eq:w_local}
    \hat W=\sum_i \hat W_{\mathrm{local},i}+\hat W_{\mathrm{non-local}}
\end{equation}
\mods{as schematically shown in Fig.~\ref{fig:local_approx}.}
With the separation given in Eq.~\eqref{eq:w_local} the RDMF can be approximated as a sum of local RDMFs and a remainder containing the non-local interactions as
\begin{equation}\label{eq:F_local_decomp}
    F^{\hat W}[\mat \rho^{(1)}]\approx \sum_i F^{\hat W_{\mathrm{local},i}}[\mat \rho^{(1)}]+F^{\hat W_{\mathrm{non-local}}}[\mat \rho^{(1)}].
\end{equation}
The RDMFs $F^{\hat W_{\mathrm{local},i}}[\mat \rho^{(1)}]$ have only a local interaction but are still functionals of the full one-particle reduced density matrix $\mat \rho^{(1)}$ of the full system. 
\mods{The non-local interactions, $\hat W_{\mathrm{non-local}}$, and the corresponding RDMF can either be similarly decomposed further into semi-local RDMFs or be evaluated approximately from approximate parametrized RDMFs~\cite{Pernal2016}, or even approximate parametrized density functionals.}

\mods{The following section shows how the ACA~\cite{PhysRevB.97.245131} can be used to evaluate a local or semi-local RDMF $F^{\hat \tilde W}[\mat \rho^{(1)}]$ by creating a smaller effective system for which the RDMF has to be evaluated.}

\subsection{\label{sec:ACA}Adaptive Cluster Approximation}
The starting point of the ACA~\cite{PhysRevB.97.245131} is a (semi-)local RDMF $F^{\hat W_\mathrm{local}}[\mat \rho^{(1)}]$ with an interaction 
\begin{equation}
    \hat W_{\mathrm{local}}=\frac{1}{2} \sum_{\alpha,\beta,\gamma,\delta \in C} U_{\alpha,\beta,\delta,\gamma} \hat c^\dagger_\alpha \hat c^\dagger_\beta \hat c_\gamma \hat c_\delta,
\end{equation}
which only includes a limited number of orbitals $N_\mathrm{int}=|C|$ that is much lower than the total number of one-particle basis states $N_\chi$ \mods{of the one-particle reduced density matrix $\mat \rho^{(1)}$} . 

The goal of the ACA is to approximate the RDMF by modifying the one-particle reduced density matrix $\mat \rho^{(1)}$ into a much smaller one-particle reduced density matrix $\mat \rho^{(1)}_{ACA}$ so that
\begin{equation}
    F_{ACA}^{\hat W_\mathrm{local}}[\mat \rho^{(1)}]=F^{\hat W_\mathrm{local}}[\mat \rho^{(1)}_{ACA}] \approx F^{\hat W_\mathrm{local}}[\mat \rho^{(1)}]
\end{equation}
and
\begin{equation}
    \frac{\partial F_{ACA}^{\hat W_\mathrm{local}}[\mat \rho^{(1)}]}{\partial \rho^{(1)}_{\beta,\alpha}}\approx \frac{\partial F^{\hat W_\mathrm{local}}[\mat \rho^{(1)}]}{\partial \rho^{(1)}_{\beta,\alpha}}.
\end{equation}
\mods{This is achieved by systematically constructing an environment by setting up a unitary transformation of the one-particle states which do not interact in $\hat W_\mathrm{local}$ so that the transformed density matrix has a banded shape. This increases the \textit{nearsightedness}~\cite{ProdanKohn} of the one-particle reduced density matrix and drastically mediates the impact of a truncation of most of the non-interacting one-particle states. If only the interacting states are kept, it is denoted as zeroth-order ACA ($\mathrm{ACA}_0$), whose the zeroth-order ACA density matrix has $N_\mathrm{int}$ one-particle states. The first-order ACA is defined by keeping all one-particle states that have non-zero elements in the transformed density matrix with interacting one-particle states. Hence, the first-order ACA density matrix has at most $2N_\mathrm{int}$ one-particle states. Thus, the solution of the RDMF in the $n$-order ACA would require at most $(n+1)N_\mathrm{int}$ one-particle states and, hence, greatly reduce the number of required qubits.}

\begin{figure}[t]
    \centering
    \includegraphics[width=0.5\textwidth]{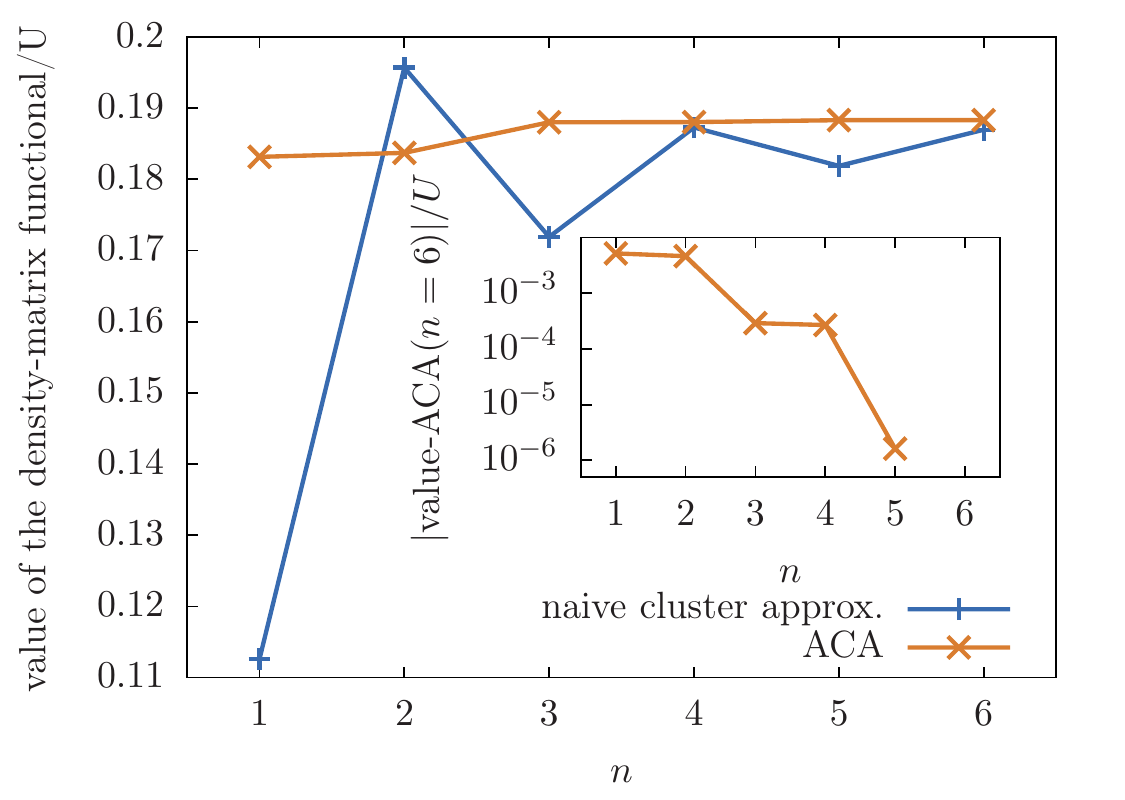}
    \caption{Convergence of the RDMF $F^{\hat W_{\mathrm{local},1}}[\mat \rho^{(1)}_{\mathrm{ACA}(n)}]$ of the ground-state density matrix for a half-filled 16-site Hubbard chain (U/t=1) for the local interaction $\hat W_{\mathrm{local},1}$ on the first site of chain: the $n$-th order ACA is compared to the value of the RDMF $F^{\hat W_{\mathrm{local},1}}[\mat \rho^{(1)}_{\mathrm{naive},n}]$ with a naive truncation, where the truncated density matrix $\mat \rho^{(1)}_{\mathrm{naive},n}$ only considers the first $n$ sites of the chain.} 
    \label{fig:aca_hubbardF}
\end{figure}

Fig.~\ref{fig:aca_hubbardF} compares the convergence of the RDMF in the $n$-th order ACA, i.e. $F^{\hat W_{\mathrm{local},1}}[\mat \rho^{(1)}_{\mathrm{ACA}(n)}]$ with the naive cluster approximation, where only the first $n$ sites of the chain are considered for the RDMF $F^{\hat W_{\mathrm{local},1}}[\mat \rho^{(1)}_{\mathrm{naive},n}]$. Although both RDMFs have the same number of one-particle states $nN_\mathrm{int}$ and hence the same computational complexity, the RDMF in the ACA converges much more smoothly and quickly with the number of one-particle states. 
The numerical results suggest that the additive error 
\begin{align}
\epsilon(n)&=\left |F^{\hat W_{\mathrm{local},1}}[\mat \rho^{(1)}_{\mathrm{ACA}(n)}]-F^{\hat W_{\mathrm{local},1}}[\mat \rho^{(1)}]\right|
\end{align}
converges like $\mathcal{O}(e^{-n})$. In other words, the required ACA-level $n$ and, hence, the qubit count is in $\mathcal{O}(\log(\epsilon^{-1}))$.

The corresponding implementation for the local approximation and the ACA for the Hubbard model is publicly available~\cite{robert_schade_2021_5749768}.


\subsection{\label{sec:RDMFT_VQE}Comparison of an RDMFT-Based Approach to Wave-Function Based Approaches}
\mods{The proposed RDMFT-based approach requires a two-level minimization procedure expressed by Eq.~\eqref{eq:EN} and Eq.~\eqref{eq:F_para}, respectively. 
This is in contrast to most wave function based approaches either on classical or quantum computers via the VQE.
The two-level nature together with the fact that constrained minimization problem are commonly more complicated to solve than unconstrained minimization procedures shows that the proposed approach is more involved than for example a conventional VQE procedure.}

\mods{The impact of the inner level, i.e., the constrained minimization of the RDMF can be mediated in practice by reusing the optimal parameters from the previous total-energy minimization step. Apart from the straightforward availability of analytical nuclear forces, the main advantage of the proposed approach lies in its parallel nature and the ability to treat much larger systems than the underlying wave function representation (either classical or quantum-computing based) would allow.}

\mods{As already noted in section~\ref{sec:local}, our formulation within the framework of RDMFT also permits to utilize parametrized one-particle reduced density matrix functionals~\cite{Pernal2016,PhysRevLett.119.063002} for some of the local or non-local RDMFs in Eq.~\eqref{eq:F_local_decomp}, which have already been shown to describe static and dynamic correlation well. Thus, precious quantum computing resources can be reserved for orbitals or spatial regions with otherwise hard to describe strong electronic correlations.}

\section{\label{sec:RDMFT_QC}Hybrid Quantum-Classical Algorithm for the RDMF}
\begin{figure}[t]
    \centering
    \includegraphics[width=0.45\textwidth]{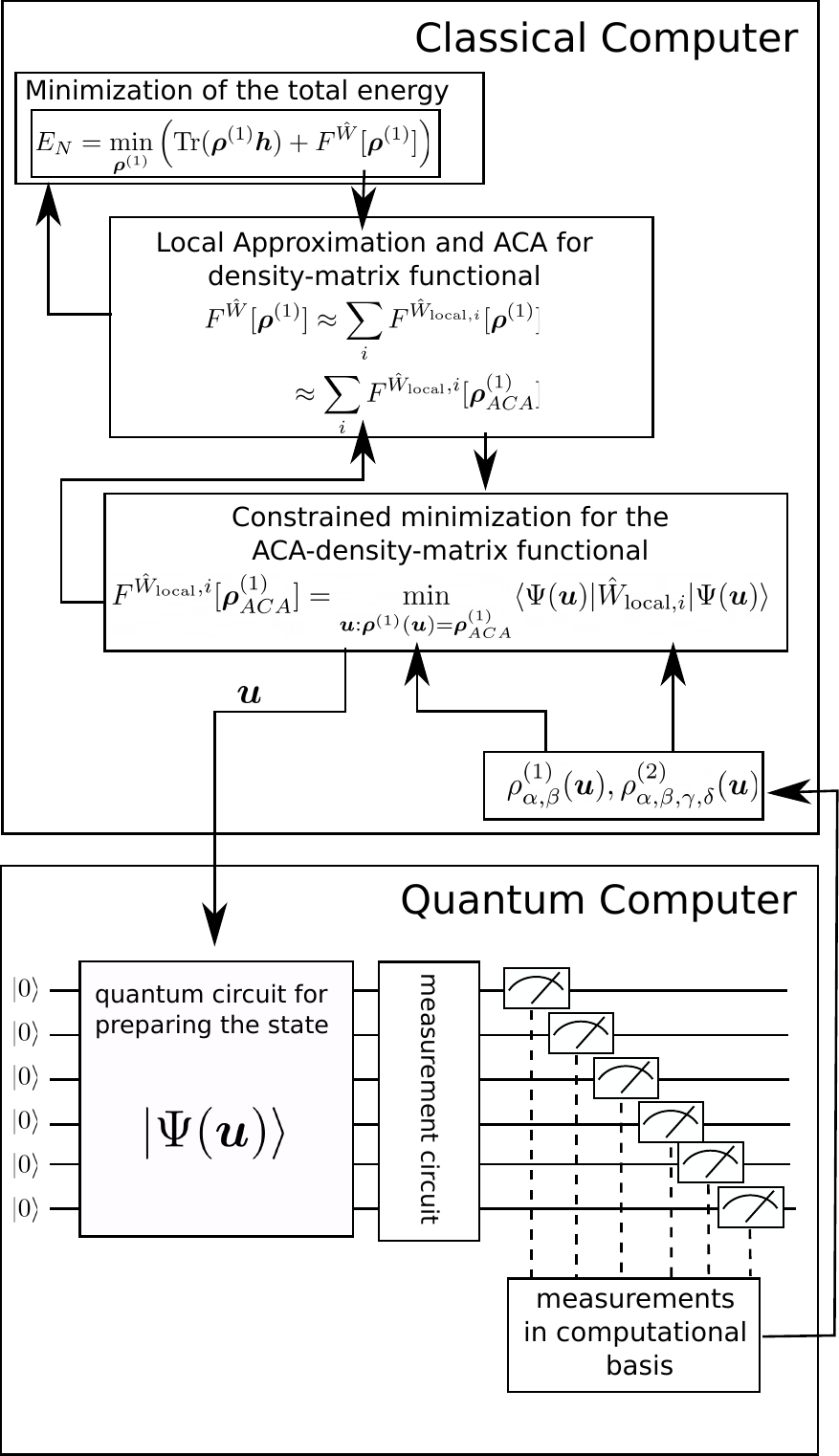}
    \caption{Schematic representation of our hybrid quantum-classical algorithm for the total energy minimization with the RDMF. The RDMF is approximated with the local approximation and the ACA so that the number of required qubits is drastically reduced in comparison to existing VQE-like approaches.}
    \label{fig:schema}
\end{figure}

We propose to evaluate the density-matrix \mods{functionals $F^{\hat{\tilde W}}[\mat \rho^{(1)}]$ in Eq.~\eqref{eq:F_local_decomp}} on quantum computers with a VQE-like approach by solving the constrained minimization problem given in Eq.~\eqref{eq:F1}. For this purpose, a parametrized ansatz $|\Psi(\mat u)\rangle$ is chosen for the fermionic many-particle wave state $|\Psi\rangle$. It is parameterized by a vector \mods{of} real parameters $\mat u$. Several different efficient parametrizations are possible and are discussed in Sec.~\ref{sec:HETS}.
A schematic of the computational approach is shown in Fig.~\ref{fig:schema}

The minimum of the parametrized constrained minimization 
\begin{equation}
\label{eq:F1u}
    F_{\mat u}^{\hat{\tilde W}}[\mat \rho^{(1)}]=\min_{\mat u: |\Psi(\mat u)\rangle \rightarrow \mat \rho^{(1)}} \langle\Psi(\mat u)|\hat {\tilde  W}|\Psi(\mat u)\rangle
\end{equation}
is an upper bound to the exact constrained minimum of Eq.~\eqref{eq:F1} if the parametrization is sufficiently flexible to allow the fulfillment of the density-matrix constraints, i.e. there exists a $\mat u$ so that
\begin{equation}
    \rho^{(1)}_{\beta,\alpha}=\langle \Psi(\mat u)|\hat c^\dagger_\alpha \hat c_\alpha|\Psi(\mat u)\rangle
\end{equation}
for the given $\mat \rho^{(1)}$.

Similar to the VQE this can be used to construct a hybrid quantum-classical algorithm, where the parameters $\mat u$ are optimized on a classical computer, but all expectation values of quantum mechanical observables in the state $|\Psi(\mat u)\rangle$ are evaluated on a quantum computer. 
Relevant observables to be evaluated on the quantum computer are the elements of the one-particle reduced density matrix
\begin{equation}
    \rho^{(1)}_{\beta,\alpha}(\mat u)=\langle \Psi(\mat u)|\hat c^\dagger_\alpha \hat c_\alpha|\Psi(\mat u)\rangle
\end{equation}
and elements of the two-particle reduced density matrix $\mat \rho^{(2)}$, i.e.
\begin{equation}
    \rho^{(2)}_{\alpha,\beta,\gamma,\delta}(\mat u)=\langle \Psi(\mat u)|\hat c^\dagger_\gamma \hat c^\dagger_\delta \hat c_\alpha \hat c_\beta|\Psi(\mat u)\rangle.
\end{equation}
Only those elements are required for which the corresponding matrix elements of the interaction Hamiltonian are non-zero.
With these quantities, the expectation value of the interaction Hamiltonian $\tilde W(\mat u)=\langle \Psi(\mat u)|\hat{\tilde W}|\Psi(\mat u)\rangle$ can be estimated. These expectation values carry the exponential complexity of the fermionic many-particle problem and can be evaluated efficiently with gate-based quantum computers. 


The practical evaluation of expectation values requires two steps. In the first step, the state $|\Psi(\mat u)\rangle$ is set up on the quantum computer. The second step is to measure the fermionic operators in the prepared state. To measure fermionic observables on gate-based quantum computers they have to be transformed to bosonic qubit-operators which in turn have to be transformed with additional gates to Pauli-z operators that can finally be measured. Details about fermionic transformations are discussed in Sec.~\ref{sec:transform}.

With the measurements of the expectation values, the constrained minimization for the RDMF in Eq.~\eqref{eq:F1u} can be written as
\begin{equation}\label{eq:F_para}
    F^{\hat{\tilde  W}}[\mat \rho^{(1)}]=\min_{\mat u: \mat \rho^{(1)}(\mat u)=\mat \rho^{(1)}} W(\mat u).
\end{equation}
To solve this constrained minimization problem we propose to use the augmented Lagrangian approach~\cite{powellmethod,Hestenes1969}, as described in Algorithm~\ref{algo:rdmf_qc}.
To formulate the problem in terms of the standard constrained optimization problem we map the complex density-matrix constraints $\mat \rho^{(1)}(\mat u)-\mat \rho^{(1)}=0$ to equality constraints $c_i$ for real and imaginary parts individually. Then the augmented Lagrangian can be written as
\begin{equation}
    L(\mat u,\mat \lambda,\mat \mu)=W(\mat u) +\sum_{i} \lambda_{i} c_i(\mat u)+\frac{1}{2}\sum_{i} \mu_i (c_i(\mat u))^2,
\end{equation}
with the objective function $W(\mat u)$, the Lagrange multipliers $\lambda_i$ and penalty parameters $\mu_i$. After setting initial penalty parameters $\mat \mu_0=\mat \mu_{\mathrm{initial}}$ and Lagrange multipliers $\mat \lambda_0=\mat \lambda_\mathrm{initial}$ the main loop of the augmented Lagrangian begins and requires in every loop iteration $k$ the solution of the auxiliary unconstrained problem 
\begin{equation}
\label{eq:unconstr}
    \mat u_k=\mathrm{argmin}_{\mat u} L(\mat u,\mat \lambda_k,\mat \mu_k).
\end{equation}
The constraint violations $c_i(\mat u_k)$ in this iteration are then used to update the Lagrange multipliers as $\lambda_{i,k+1}=\lambda_{i,k}+\mu_{i,k}c_i(\mat u_k)$ (first-order multiplier update).
\mods{Additionally, the penalty parameters $\mat \mu_{k}$ can be modified for example with a constant factor $\beta>1$, i.e., $\mat \mu_{k+1}=\beta \mat \mu_k$ before the algorithm proceeds to the next iteration $k \rightarrow k+1$.}
Thus, for the solution of the constrained minimization only unconstrained problems in Eq.~\eqref{eq:unconstr} have to be solved. Existing techniques from noisy unconstrained minimization such as the simultaneous perturbation stochastic approximation (SPSA)~\cite{119632,4789489} or related noise-resistant algorithms can be used.

The augmented Lagrangian approach as a method for constrained minimization has the advantage that no derivatives of the objective function or constraints are required and that exact constraint satisfaction is not required in every iteration. A major advantage compared to the use of penalty methods is that due to the presence of the Lagrange multipliers the penalty parameters do not have to be increased to infinity which drastically reduces the problem of ill-conditioning. The cost to be paid for this advantage is the necessity for multiple solutions of the unconstrained subproblems in Eq.~\eqref{eq:unconstr} that, however, can be mediated by a warm-start strategy. The augmented Lagrangian approach for the RDMF usually converges in between five and 10 outer iterations of the augmented Lagrangian~\cite{schadephd}.

A further important property of the augmented Lagrangian approach is the fact that the Lagrange multipliers $\lambda_i$ correspond to the derivatives of the RDMF with respect to the elements of the one-particle reduced density matrix which is required for an efficient minimization of the total energy in Eq.~\eqref{eq:EN}.
\begin{algorithm}[H]
\caption{RDMF with Augmented Lagrangian}\label{algo:rdmf_qc}
\begin{algorithmic}
\State set initial parameters $\mat u$, initial penalties $\mat \mu$, initial Lagrange multipliers $\mat \lambda$
\While{not converged}
    \State solve unconstrained problem $\min_{\mat u} L(\mat u,\mat \lambda,\mat \mu)$
    \State $\mat \lambda \gets \mat \lambda+\mat \mu \mat c(\mat u)$ \Comment{multiplier update}
    \State $\mat \mu \gets f(\mat \mu,\mat c(\mat u))$ \Comment{penalty update}
\EndWhile
\end{algorithmic}
\end{algorithm}

\section{\label{sec:qcdepth}Reduction of the Number of Qubit Operations}
\subsection{\label{sec:transform}Fermionic Transformations}
A well-known way to map fermionic operators to bosonic operators is the Jordan-Wigner transformation~\cite{Jordan1993}. It maps each one-particle state to a qubit, but the number of required additional qubit operations scales linearly with the number of qubits. The parity transformation~\cite{doi:10.1063/1.4768229} is equivalent to the Jordan-Wigner transformation in this regard. 
A promising alternative is the Bravyi-Kitaev transformation~\cite{BRAVYI2002210,doi:10.1063/1.4768229}, which in contrast to the Jordan-Wigner transformation requires only a logarithmically scaling number of additional qubit operations and becomes advantageous for large problems.

\subsection{\label{sec:HETS}Hardware-Efficient Trial States}
\mods{As a variational ansatz for the hybrid-classical algorithm, described in Sec.~\ref{sec:RDMFT_QC}, we have implemented hardware-efficient trial states~\cite{Kandala2017}. For the latter, the state is parametrized by quantum gates that are natively supported by the specific quantum device at hand. Only the fermionic measurement operators are transformed for example with transformations such as the Jordan-Wigner transformation to bosonic qubit operators. This leads to short quantum programs for the state preparation.
By contrast, some other popular trial states, such as the unitary coupled-cluster ansatz~\cite{OMalley2016, Shen2017}, where the trial state is formulated with fermionic operators that are then transformed to qubit operators, tend to produce longer quantum programs for the state preparation. Our RDMFT-based approach proposed in Sec.~\ref{sec:RDMFT} is independent of the ansatz for the trial state.}

\begin{figure}[t]
    \centering
    \includegraphics[width=0.48\textwidth]{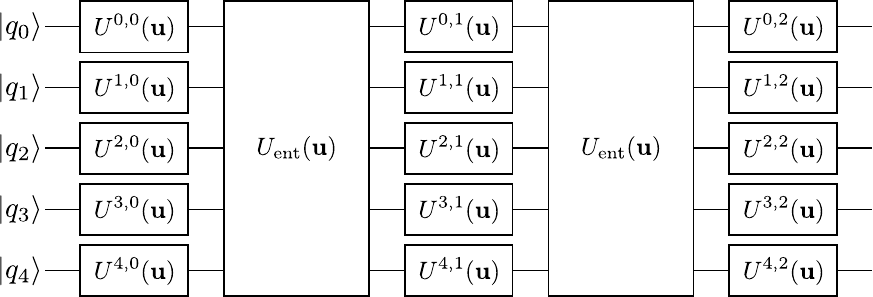}
    \caption{Quantum circuit representation of the hardware-efficient trial state composed of single-qubit Euler rotations $U^{q,i}(\mat u)$ and parametrized quantum entanglers $U_{\text{ent}}(\mat u)$.} 
    \label{fig:HETS}
\end{figure}

Specifically, we parametrize the N-qubit wave function as
\begin{equation}
|\Psi(\mat u)\rangle=\prod_{i=1}^{d}\left[\left(\prod_{q=0}^{N-1} U^{q, d-i+1}(\mat u)\right) U_{\text {ent }}\right] \prod_{q=0}^{N-1} U^{q, 0}(\mat u)|\mat 0\rangle \label{eq:HETS}
\end{equation}
in which the index $i$ enumerates $d$ repetitive blocks of single-qubit Euler rotations $U^{q, i}(\mat u)$ followed by entanglers $U_{\text {ent}}$, which are composed of a sequence of two-qubit gates to create an entanglement between the qubits (Fig.~\ref{fig:HETS}). Crucially, we choose the $U^{q, i}(\mat u)$ as a sequence of single-qubit gates in accordance with the features of the quantum computer that is available to us.

In the form of Eq.~\eqref{eq:HETS}, the blocked structure of the conventional hardware-efficient trial states leads to an extendability issue: parameters that were optimized for a 
trial state with a given number of blocks can not be used as a starting point for a trial state with more blocks due to the presence of the entanglers. However, one can work around 
this caveat by replacing the conventional entanglers $U_{\text{ent}}$ by parametrized entanglers $U_{\text{ent}}(\mat u)$ that for a certain choice of the parameters are identical 
to the identity-gate. In this case, a shallow initial trial state can be readily extended with these "disabled" entanglers, which will then be tuned away from identity as part of 
the constrained minimization.
A simple choice for the parametrized entanglers is the controlled rotations, i.e. CRX or CRZ, instead of the basic controlled Pauli gates  CNOT or CZ. Note that one drawback 
of parametrizing the entanglers is the doubling of required basic two-qubit gates for the implementation of $U_{\text{ent}}(\mat u)$  due to the decomposition of a controlled 
rotation into two controlled Pauli gates and additional single-qubit gates.




\subsection{\label{sec:flex}Increasing the Flexibility of the Trial States}
We are not aware of any method that can answer the question if a chosen hardware-efficient trial state can represent the ground state or any other desired state of a system in polynomial time by only using classical computations. Thus, any additional degrees of freedom in the hardware-efficient trial state that do not affect the computation on a quantum computer and only consume polynomial time are useful. 
Within the ACA, there are $N_\mathrm{int}$ interacting one-particle states and $nN_\mathrm{int}$ non-interacting one-particle states. We therefore propose here to introduce an additional degree of freedom by means of an unitary transformation of the non-interacting states that is optimized in every step of the solution of the unconstrained subproblems to minimize the augmented Lagrangian. 
This does not increase the number of observables to be measured, because the interactions remain local on the $N_\mathrm{int}$ sites. The computational effort for this additional step is polynomial in $(n+1)N_\mathrm{int}$. The additional freedom through the rotation of part of the one-particle basis makes a given parametrized hardware-efficient trial state more flexible in the sense that it can represent fermionic states that were not representable without the additional unitary transform.

\section{\label{sec:programcount}Reduction of the Number of Quantum Programs}
At variance to the VQE, where only the expectation value of the Hamiltonian $\hat H$ is required, the proposed evaluation of the RDMF on quantum computers requires the individual estimation of all elements of the one-particle reduced density matrix, i.e. $\rho^{(1)}_{\alpha,\beta}$ and the expectation value of the interaction Hamiltonian $\hat W$. 

\subsection{Measurements for the Interaction Hamiltonian}
Measurements for the expectation value of the interaction $\langle \hat W\rangle$ or the local interaction $\langle \hat W_{\mathrm{local},i}\rangle$ require in the worst case $\mathcal{O}(N_\chi^4)$ quantum programs and $\mathcal{O}(N_{\mathrm{int},i}^4)$ quantum programs, respectively.
The measurement of the expectation value of the (local) interaction Hamiltonian is very similar to the measurement of the total energy in the VQE because the interaction is contained 
in the total energy. Thus, the same techniques developed for the VQE can be applied here for the interaction energy, e.g. the approach proposed by Izmaylov et al.~\cite{doi:10.1021/acs.jctc.9b00791} brings the scaling down to $\mathcal{O}(N_\chi^3)$ and $\mathcal{O}(N_{\mathrm{int},i}^3)$, respectively.
However, in this work we will instead focus on the measurement of the individual elements of the one-particle reduced density matrix, which is qualitatively different from the measurement of the total energy or interaction energy.

\subsection{Measurements for the One-Particle Reduced Density Matrix}
The number of measurements to evaluate an individual element of the one-particle reduced density matrix is independent of the system size for the common fermionic transformations. Thus, the number of measurements for each evaluation of the one-particle reduced density matrix scales quadratically with the number of one-particle basis states. Consequently, without the ACA of the RDMF, the number of measurements would scale quadratically with the overall system size in terms of one-particle basis states.

With the ACA, the number only scales as 
\begin{equation}
\mathcal{O}\left (\sum_{i=1}^{N_\mathrm{local}} (n+1)^2N_{\mathrm{int},i}^2\right),
\end{equation}
where $N_\mathrm{local}$ is the number of local RDMFs, $n$ the chosen order of the ACA, $N_{\mathrm{int},i}$ the number of interacting one-particle states in the $i$-th local interaction $\hat W_{\mathrm{local},i}$. The number of local RDMFs is linear in terms of the overall system size. 
Provided some order $n$ of the ACA is sufficient for a proper description of the system, then the overall number of measurements for the one-particle reduced density matrices scale linearly with the system size because the number of measurements for each local RDMF is independent of the overall system size.
Thus, this linear-scaling behavior of the ACA drastically reduces the qubit requirements, as well as the overall measurement count.

\subsubsection{Combination of Measurements}
The number of quantum programs can be further reduced by combining commuting observables~\cite{McClean_2016,Kandala2017,gokhale2019minimizing} instead of measuring only a single observable per quantum program execution. We will show here that all elements of the one-particle reduced density matrix can be measured with only $\mathcal{O}(N_\chi)$ quantum programs instead of $\mathcal{O}(N_\chi^2)$, where $N_\chi$ is the number of one-particle states.
The fermionic transformation turns the elements of the one-particle reduced density matrix $\hat \rho_{\beta,\alpha}=\hat c_\alpha^\dagger \hat c_\beta$ into weighted sums of Pauli strings, e.g. $\hat c^\dagger_1 \hat c_1 \rightarrow 1/2 \mathbbm{1} \otimes ...+ 1/2 \hat Z \otimes \mathbbm{1} \otimes ... $ in the Jordan-Wigner transformation. 
Pairwise commuting Pauli strings can be measured simultaneously, i.e. in a single quantum program.
We define three different levels of commutation
\begin{itemize}
    \item DISJOINT: Disjoint-qubit commutativity means that the Pauli operators in the two Pauli strings act on different qubits, e.g. $\hat Z \otimes \mathbbm{1}$ and $\mathbbm{1} \otimes \hat Z$ commute, but not $\hat Z \otimes \mathbbm{1}$ and $\hat Z \otimes \hat Z$.
    \item QWC: Qubit-wise commutativity is satisfied if each Pauli operator in the first Pauli string commutes with the Pauli operator on the same qubit of the second Pauli string, e.g. $\hat Z \otimes \mathbbm{1}$ and $\hat Z \otimes \hat Z$ commute, but not $\hat Z \otimes \mathbbm{1}$ and $\hat X \otimes \hat Z$, respectively.
    \item GC: general commutativity between two Pauli strings, e.g. $\hat Z \otimes \hat X\otimes \hat X$ and  $\mathbbm{1} \otimes \hat Y \otimes \hat Z$ commute.
\end{itemize}
The employed commutation level can be used to define a graph where the Pauli strings that have to be measured are the nodes and the two nodes are connected by an edge if they commute at the selected commutation level. The problem of finding the lowest possible number of disjoint groups of pairwise commuting Pauli strings, i.e. sets of Pauli strings that can be measured simultaneously is equivalent to the minimum clique cover problem. The minimum clique cover problem can be solved with heuristic graph coloring algorithms. 

All Pauli strings in a clique are then measured with one quantum program and the total number of quantum programs is given by the number of cliques. Even though the minimum clique cover
problem is NP-hard~\cite{kar72}, greedy algorithms have been shown to give good approximate results in polynomial time. We use here the algorithms implemented in the NetworkX-library~\cite{SciPyProceedings_11}.
For the remainder of the section, we use the Jordan-Wigner transformation and assume an even number of one-particle states. Other fermionic transformations give similar results.

\begin{figure}[t]
    \centering
    \includegraphics[width=0.5\textwidth]{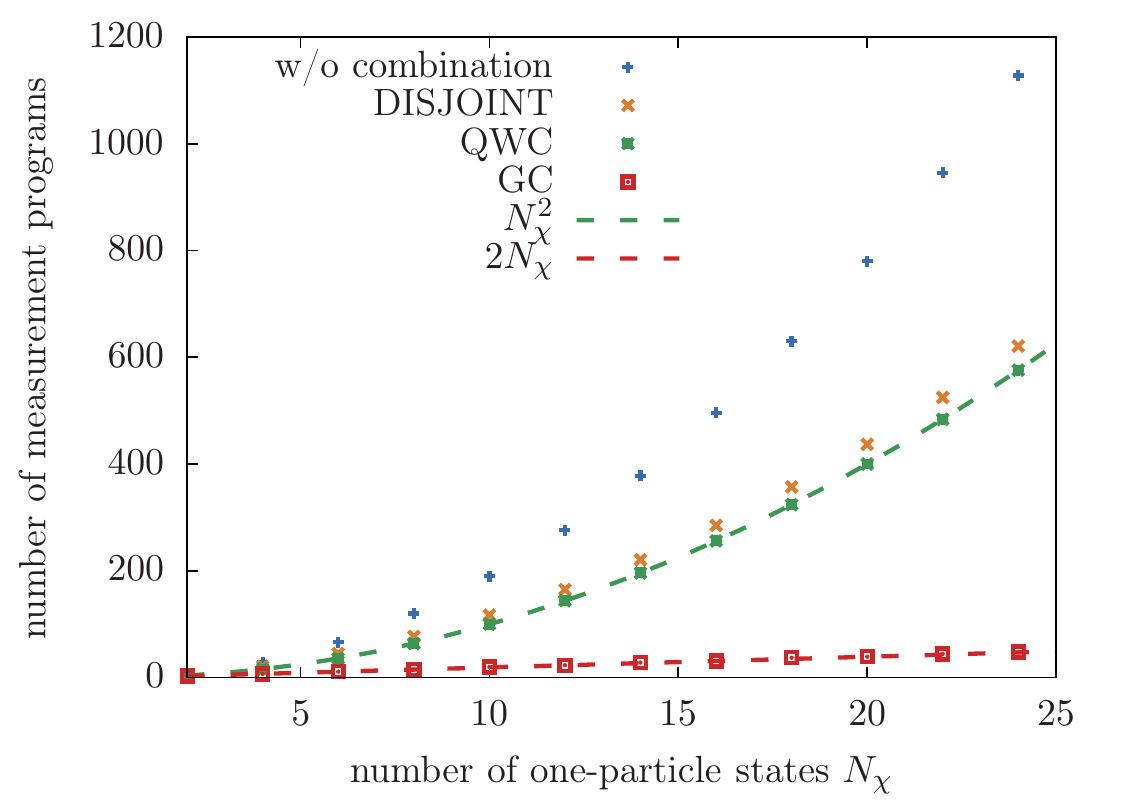}
    \caption{The number of quantum programs for the measurement of the one-particle reduced density matrix for a given number of one-particle states at different levels of commutativity of Pauli strings: without commutativity, Pauli strings acting on disjoint qubits (DISJOINT), qubit-wise commutativity (QWC) and general commutativity (GC), respectively. The Jordan-Wigner transformation has been used and the implied minimum clique cover problem has been solved with polynomial-time algorithms from NetworkX.} 
    \label{fig:commutativity}
\end{figure}

Fig.~\ref{fig:commutativity} shows the dependence of the number of measurement programs on the number of one-particle states $N_\chi$ for the measurement of the one-particle reduced density matrix for the three commutativity levels.
In total $2N_\chi^2-N_\chi$ Pauli strings have to be measured for the elements of the one-particle reduced density matrix. When using qubit-wise commutativity, the number is reduced to roughly $N_\chi^2$, but still scales quadratically. 
Using the general commutativity of Pauli strings, the number of programs is reduced to approximately $2N_\chi$ because with each quantum program about $N_\chi$ Pauli strings can be measured.

\paragraph{Construction of Measurement Programs}
When measuring just a single Pauli string, the measurement circuits are straightforward to set up by first transforming the x- and y-Pauli operators to Pauli-z operators with single-qubit gates and then reducing the multi-qubit z-measurements to single-qubit z-measurements with two-qubit gates. The freedom in the placement of the two-qubit gates can be used to optimize for the coupling topology of the given quantum computer.

When measuring multiple Pauli strings per quantum program, we follow the general idea of the construction given by Gokhale et al.~\cite{gokhale2019minimizing} based on the stabilizer formalism~\cite{1997PhDT232G,PhysRevA.70.052328}.  However, we have improved on the formulation and implementation of the construction in several aspects: we have implemented it purely over GF(2), we have added the missing phase row, our construction also works if the set of Pauli strings is not complete or linearly independent, and we have identified freedom in some aspects of the construction to optimize for a given quantum computer.
Because of our modifications in the underlying algorithm, we present the construction here as a whole.

The construction utilizes the stabilizer matrix-formalism which writes a set of $N$ commuting Pauli strings that act on $N_q$ qubits as a stabilizer matrix $S\in \mathrm{GF}(2)^{2N_q+1 \times N}$, i.e. 
\begin{align}
    S&=\begin{bmatrix}
    Z_{1,1} & Z_{1,2} & ... & Z_{1,N} \\
    \vdots & \vdots & ... & \\
    Z_{N_q,1} & Z_{N_q,2} & ... & Z_{N_q,N} \\ \hline
    X_{1,1} & X_{1,2} & ... & X_{1,N} \\
    \vdots & \vdots & ... &\\
    X_{N_q,1} & X_{N_q,2} & ... & X_{N_q,N} \\ \hline
    r_{1} & r_{2} & ... & r_{N}
    \end{bmatrix}.
\end{align}
The elements are defined as follows: 
\begin{enumerate}
    \item $Z_{i,j}=1$ iff the \mods{$j$}-th Pauli-string has the Pauli-z or Pauli-y acting on the \mods{$i$}-th qubit.
    \item $X_{i,j}=1$ iff the \mods{$j$}-th Pauli-string has the Pauli-x or Pauli-y acting on the \mods{$i$}-th qubit.
    \item \mods{$r_j$ defines that the $j$-th single-qubit measurement needs to be multiplied with $(-1)^{r_j}$}.
    \item Otherwise all elements are zero.
\end{enumerate}
For example the Pauli strings $\hat X\otimes \hat Z\otimes \hat Y\otimes \mathbbm{1}$, $\hat Y\otimes \hat Z\otimes \hat X\otimes \mathbbm{1}$, $\mathbbm{1}\otimes \hat X\otimes \hat Z \otimes \hat Y$, and $\mathbbm{1}\otimes \hat Y\otimes \hat Z \otimes \hat X$ correspond to the stabilizer matrix
\begin{align}
\label{eq:S_example}
    S=\begin{bmatrix}
    0 & 1 & 0 & 0 \\
    1 & 1 & 0 & 1 \\
    1 & 0 & 1 & 1 \\
    0 & 0 & 1 & 0 \\ \hline
    1 & 1 & 0 & 0 \\ 
    0 & 0 & 1 & 1 \\
    1 & 1 & 0 & 0 \\
    0 & 0 & 1 & 1 \\ \hline
    0 & 0 & 0 & 0 \\
    \end{bmatrix},
\end{align}
where $r_i=0$ has been chosen as a convention.
The action of quantum gates now corresponds to changes in the stabilizer matrix~\cite{PhysRevA.70.052328}:
\begin{itemize}
    \item CNOT(c,t): $\forall i \in {1,...,N}$ do $r_i \rightarrow r_i + Z_{t,i}X_{c,i}(1+X_{t,i}+Z_{c,i})$, $X_{t,i}\rightarrow X_{t,i}+X_{c,i}$, $Z_{c,i}\rightarrow Z_{c,i}+Z_{t,i}$.
    \item Hadamard gate H(q): $\forall i \in {1,...,N}$ do $r_i \rightarrow r_i + X_{q,i}Z_{q,i}$ and swap $X_{q,i}$ and  $Z_{q,i}$.
    \item Phase gate S(q): $\forall i \in {1,...,N}$ do $r_i \rightarrow r_i + X_{q,i}Z_{q,i}$, $Z_{q,i}\rightarrow Z_{q,i}+X_{q,i}$.
\end{itemize}
All operations are performed in GF(2). The inclusion of the phase row $r_1,...,r_N$ guarantees that the sign of the measurements is correct.

The algorithm for the construction of the measurement circuits is described in Algorithm~\ref{algo:constr1}. To simplify the notation we define the first $N_q$ rows of the stabilizer matrix as matrix $S_Z$, the following $N_q$ rows as the matrix $S_X$, and the last row as the phase row.
\begin{algorithm}[H]
\caption{Construction of Measurement Programs}\label{algo:constr1}
\begin{algorithmic}
\State \textbf{Input:} $N$ commuting Pauli strings $P_1,...,P_N$
\State \textbf{Output:} measurement circuit consisting of the gates that have been applied to $S$
\State $S \gets$ stabilizer matrix of $P_1,...,P_N$ with $r_i=0$  \Comment{prep.}
\State $$R_{ZX} \gets rk\left (\begin{bmatrix}S_Z \\ S_X \end{bmatrix}\right)$$
\While{$rk(S_X) < R_{ZX}$} \Comment{rank max.}
    \State choose some qubit $q$ to apply Hadamard to
    \State $S \gets H(q)S$
\EndWhile
\State \mods{$P,L,U \gets PLU$} decomposition of $S_{ZX}$ \Comment{PLU-decomp.}
\State $T \gets $ transpositions in $P$
\For{each transposition $i \rightarrow j$ in $T$} \Comment{permutation}
    \State $S \gets \mathrm{SWAP}(i,j)S$
\EndFor
\State $L^{-1} \gets \mathrm{inv}(L)$
\For{each $(i,j), i\neq j$ with \mods{$L^{-1}_{i,j}=1$}} \Comment{row reduc.}
    \State $S \gets \mathrm{CNOT}(i,j)S$
\EndFor

\If{$S_X$ is not diagonal} \Comment{diag. reduc.}
    \State reduce $S_X$ to diagonal form with CNOTs
\EndIf
\State reduce $S_Z$ to zero with phase gates and CNOTs
\For{each qubit $q$} \Comment{X-Z-flip}
    \State $S \gets \mathrm{H}(q)S$
\EndFor
\For{each qubit $q$} \Comment{sign-step}
    \If{$r_q=1$}
        \State $S \gets \mathrm{Y}(q)S$
    \EndIf
\EndFor
\end{algorithmic}
\end{algorithm}
Specifically, the algorithm performs the following steps:
\begin{enumerate}
    \item maximizing the rank of the $S_X$ matrix (rank max.),
    \item transforming the $S_X$ to upper triangular form (PLU-decomp. and row reduc.),
    \item reducing $S_X$ to diagonal form (diag. reduc.),
    \item transforming the $S_Z$ to a zero matrix (Z-reduction),
    \item exchanging the elements of the $S_Z$ and $S_z$ matrix so that $S_X=0$ and $S_Z$ diagonal (X-Z-flip)
    \item and using the information in the phase row to get the correct signs in the measurements (sign-step).
\end{enumerate}
The result is a stabilizer matrix, where $S_Z$ is diagonal, $S_X=0$ and $r_i=0$, which means that the Pauli strings $P_i$ can be measured as single-qubit measurements at the qubits $q$, where $S_{Z,q,i}=1$.

The initial step is the construction of the stabilizer matrix $S$ from the set $\{P_i\}$ of $N$ Pauli strings. It should be noted that all algebra has to be performed over GF(2) 
instead of $\mathbb{R}$. The initial rank of the matrix $\begin{bmatrix}S_Z & S_X \end{bmatrix}^T$ represents the maximal rank of the $S_X$-matrix that can be reached 
by exchanging rows of the $S_X$ and $S_Z$ matrix with Hadamard gates.
Next, these Hadamard gates are applied to qubits such that the rank of $S_X$ becomes equal to the initial rank of $\begin{bmatrix}S_Z & S_X \end{bmatrix}^T$. The qubits where the
Hadamard gates have to be applied can be found in polynomial time with the simple Algorithm~\ref{algo:HX}.
\begin{algorithm}[H]
\caption{Maximizing the rank of $S_X$ with Hadamards}\label{algo:HX}
\begin{algorithmic}
\State \textbf{Input:} stabilizer matrix S
\State \textbf{Output:} set $H$ of qubits, where Hadamards need to be applied so that $S_X$ has maximal rank
\State $$R_{ZX} \gets rk\left (\begin{bmatrix}S_Z \\ S_X \end{bmatrix}\right)$$
\State $H \gets \emptyset$
\While{$rk(S_X)< R_{ZX}$}
    \For{each qubit $q$ not yet in $H$}
        \If{$rk((H(q)S)_X)>rk(S_X)$}
            \State $S \gets H(q)S$
            \State $H \gets H \cup q$
        \EndIf
    \EndFor
    
\EndWhile
\State 
\end{algorithmic}
\end{algorithm}

\begin{figure}[t]
    \centering
    \includegraphics[width=0.5\textwidth]{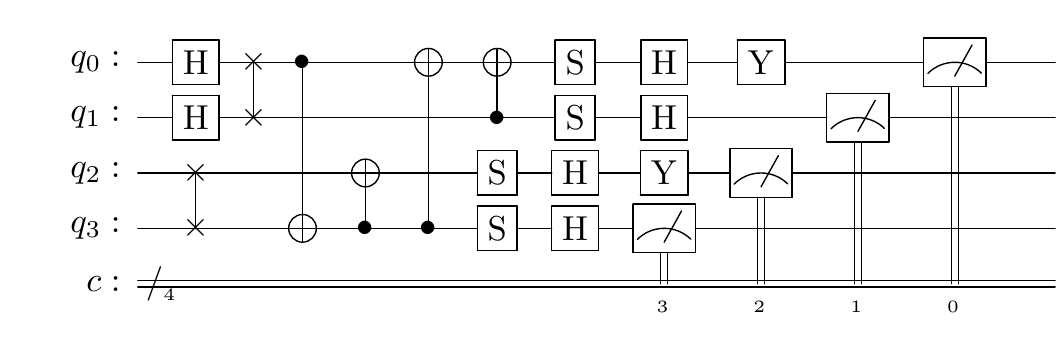}
    \caption{Measurement circuit constructed with Algorithm~\ref{algo:constr1} for the group of Pauli strings $\hat X\otimes \hat Z\otimes \hat Y\otimes \mathbbm{1}$, $\hat Y\otimes \hat Z\otimes \hat X\otimes \mathbbm{1}$, $\mathbbm{1}\otimes \hat X\otimes \hat Z \otimes \hat Y$ and $\mathbbm{1} \otimes \hat Y\otimes \hat Z \otimes \hat X$, respectively.}
    \label{fig:example_circ_unopt}
\end{figure}

For the example given in Eq.~\eqref{eq:S_example} the ranks are $rk(\begin{bmatrix}S_Z & S_X \end{bmatrix}^T)=4$ and $rk(S_X)=2$. Applying Hadamard gates at the first and second qubit results in the stabilizer matrix
\begin{align}
\label{eq:S_example_H}
    S_{\mathrm{rank\ max.}}&=\begin{bmatrix}
    1 & 1 & 0 & 0 \\
    0 & 0 & 1 & 1 \\
    1 & 0 & 1 & 1 \\
    0 & 0 & 1 & 0 \\ \hline
    0 & 1 & 0 & 0 \\ 
    1 & 1 & 0 & 1 \\
    1 & 1 & 0 & 0 \\
    0 & 0 & 1 & 1 \\ \hline
    0 & 1 & 0 & 1 \\
    \end{bmatrix}
\end{align}
with $rk(S_X)=4$. The \mods{PLU}-decomposition of $S_{\mathrm{rank\ max.},X}$ in the exemplary case is
\mods{\begin{align}
    S_{\mathrm{rank\ max.},X}&=\begin{pmatrix} 0 & 1 & 0 & 0 \\ 1 & 0 & 0 & 0 \\ 0& 0 & 0 & 1 \\ 0 & 0 & 1 & 0 \end{pmatrix}
    \begin{pmatrix}1 & 0 & 0 & 0 \\ 0 & 1 & 0 & 0 \\ 0 & 0 & 1 & 0 \\ 1 & 0 & 0 & 1 \end{pmatrix}\begin{pmatrix}
    1 & 1 & 0 & 1 \\ 0 & 1 & 0 & 0 \\ 0 & 0 & 1 &1 \\ 0 & 0 & 0 &1\end{pmatrix}
\end{align}}
and, hence, the stabilizer matrix after permutation with SWAP(0,1) and SWAP(2,3) is
\begin{align}
    S_{\mathrm{perm.}}&=\begin{bmatrix}
    0 & 0 & 1 & 1 \\
    1 & 1 & 0 & 0 \\
    0 & 0 & 1 & 0 \\
    1 & 0 & 1 & 1 \\ \hline
    1 & 1 & 0 & 1 \\ 
    0 & 1 & 0 & 0 \\
    0 & 0 & 1 & 1 \\
    1 & 0 & 0 & 0 \\ \hline
    0 & 1 & 0 & 1 \\
    \end{bmatrix}.
\end{align}
The row reduction with $L^{-1}$, i.e. CNOT(0,3) then reduces the stabilizer matrix to
\begin{align}
    S_{\mathrm{row\ red.}}&=\begin{bmatrix}
    1 & 0 & 0 & 0 \\
    1 & 1 & 0 & 0 \\
    0 & 0 & 1 & 0 \\
    1 & 0 & 1 & 1 \\ \hline
    1 & 1 & 0 & 1 \\ 
    0 & 1 & 0 & 0 \\
    0 & 0 & 1 & 1 \\
    0 & 0 & 0 & 1 \\ \hline
    0 & 1 & 0 & 1 \\
    \end{bmatrix}
\end{align}
and with the diagonal reduction, i.e. CNOT(3,2), CNOT(3,1) and CNOT(1,0) finally to
\begin{align}
    S_{\mathrm{diag.\ red.}}&=\begin{bmatrix}
    1 & 0 & 0 & 0 \\
    0 & 1 & 0 & 0 \\
    0 & 0 & 1 & 0 \\
    0 & 0 & 0 & 1 \\ \hline
    1 & 0 & 0 & 0 \\ 
    0 & 1 & 0 & 0 \\
    0 & 0 & 1 & 0 \\
    0 & 0 & 0 & 1 \\ \hline
    0 & 1 & 0 & 1 \\
    \end{bmatrix},
\end{align}
where $S_X=\mathbbm{1}$. With phase gates at all qubits, the $S_Z$ matrix can be reduced to a zero matrix and with Hadamard gates at all qubits the $S_Z$ and $S_X$ matrix are flipped to obtain 
\begin{align}
    S_{\mathrm{X-Z-flip}}&=\begin{bmatrix}
    1 & 0 & 0 & 0 \\
    0 & 1 & 0 & 0 \\
    0 & 0 & 1 & 0 \\
    0 & 0 & 0 & 1 \\ \hline
    0 & 0 & 0 & 0 \\ 
    0 & 0 & 0 & 0 \\
    0 & 0 & 0 & 0 \\
    0 & 0 & 0 & 0 \\ \hline
    0 & 1 & 0 & 1 \\
    \end{bmatrix}.
\end{align}
The final step is to apply Y gates at the second and fourth qubits to obtain the proper signs of the measurements.
The resulting measurement circuit for this example is shown in Fig.~\ref{fig:example_circ_unopt}.

\begin{figure}[t]
    \centering
    \includegraphics[width=0.5\textwidth]{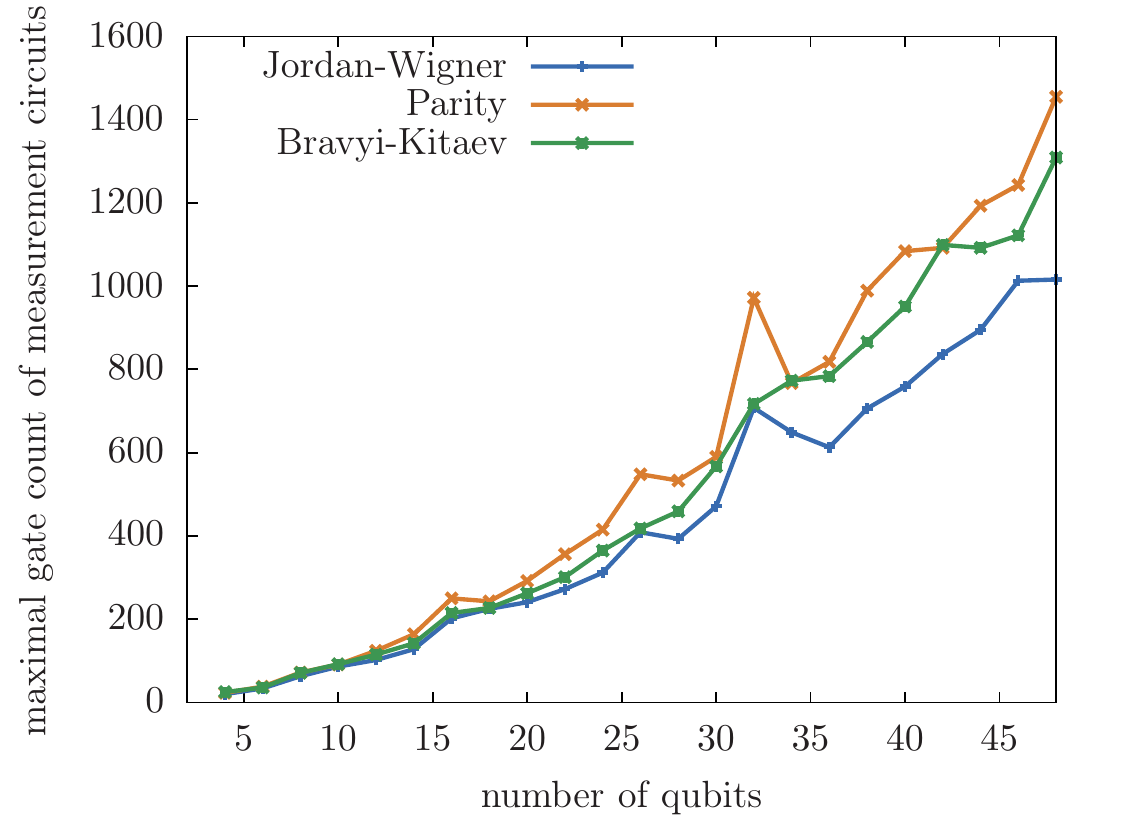}
    \caption{Maximal gate count of the constructed measurement circuits for all elements of the one-particle reduced density matrix for a given number of qubits.} \label{fig:max_gate_unopt}
\end{figure}

\paragraph{Optimization of Measurement Programs}
The naive application of the construction of measurement programs can, however, leads to an $\mathcal{O}(N^2)$ gate count in the worst case~\cite{gokhale2019minimizing}.  
Therefore, we propose here to use the degrees of freedom in the construction to optimize the measurement circuits for a given quantum computer.
One important degree of freedom is the ordering of the measurement qubits, i.e. what single-qubit measurement is measured at which qubit. The depth and the gate count of the constructed measurement circuits is sensitive to the order. Because the number of possible permutations scales exponentially with the number of qubits, an efficient heuristic for the order is required. Fixing the order so that there are no permutations in the \mods{PLU}-decomposition is an example of a simple heuristic. Fig.~\ref{fig:max_gate_unopt} shows the maximal gate count of the constructed measurement circuits for all elements of the one-particle reduced density matrix when using the mentioned simple heuristic and different fermionic transformations. Interestingly, the maximal gate count in a measurement circuit only grows roughly linear with the qubit count. 

The implementation for the presented algorithm based on Qiskit~\cite{Qiskit} is publicly available~\cite{robert_schade_2021_5749768}.
Optimizations that explore different heuristics and also consider the set of supported gates as well as the coupling topology of the given quantum computer will be addressed in future work.

\section{\label{sec:results}Results for a Model System}
\subsection{\label{sec:results_model}Definition of the Model}
The Hubbard model~\cite{gutzwiller63_prl10_159,hubbard63_prsla276_238,kanamori63_progtheorphys30_275} has proven to be an extremely valuable model system for materials with strong electronic correlations. Conventional DFT fails to qualitatively describe these strong electronic correlations and the resulting phenomena like metal-insulator transitions. Thus, alternative correlated approaches are required.

We use here a \mods{$L=8$}-site half-filled Hubbard chain with the Hamiltonian
\begin{equation}
    \hat H=-t\sum_{\sigma \in \{\uparrow,\downarrow\}}\sum_{i=1}^{L-1}\hat c_{i,\sigma}^\dagger \hat c_{i+1,\sigma}+U\sum_{i=1}^L \hat n_ {i,\uparrow} \hat n_{i,\downarrow},
\end{equation}
where $\hat n_{i,\sigma}=\hat c_{i,\sigma}^\dagger \hat c_{i,\sigma}$ is the particle number operator for the spin channel $\sigma$ on site $i$, $t$ is the hopping parameter and $U\geq 0$ the interaction operator.
The first sum represents the kinetic energy of electrons hopping between sites and the second sum covers the electron-electron interaction of two electrons on the same site. Hence, the interaction operator is
\begin{equation}
    \hat W=U\sum_{i=1}^L \hat n_ {i,\uparrow} \hat n_{i,\downarrow}.
\end{equation}
The half-filled Hubbard chain can be viewed as a model for a chain of Hydrogen atoms, where the interaction strength $U/t$ is analogous to the distance between the atoms.

We focus on the evaluation of the RDMF $F^{\hat W}[\mat \rho^{(1)}]$ for a given one-particle reduced density matrix $\mat \rho^{(1)}$. This result is then used in the minimization of the total energy in Eq.~\eqref{eq:EN}.
As a physical example case we use the one-particle reduced density matrix $\mat \rho^{(1)}_0$ of the ground-state of the chosen model. The ground state of the $L$-site half-filled Hubbard chain can be obtained efficiently for arbitrary interaction strengths with MPS-DMRG techniques. We use here the MPS implementation in ITensor~\cite{itensor} through DMRGPY~\cite{DMRGPY}.
This defines the RDMF $F^{\hat W}[\mat \rho^{(1)}_0]$ to be calculated. 

\subsection{\label{sec:results_local_aca}Local Approximation and ACA}
Now the pipeline proposed in Sec.~\ref{sec:RDMFT}, i.e. the local approximation and the ACA can be applied:
\begin{subequations}
\begin{align}
    F^{\hat W}[\mat \rho^{(1)}_0]&\approx \sum_{i=1}^L F^{U\hat n_ {i,\uparrow} \hat n_{i,\downarrow}}[\mat \rho^{(1)}_0]\\
    &\approx \sum_{i=1}^L F_{ACA(n)}^{U\hat n_ {i,\uparrow} \hat n_{i,\downarrow}}[\mat \rho^{(1)}_0]\\
    &=\sum_{i=1}^L F^{U\hat n_{i,\uparrow} \hat n_{i,\downarrow}}[\mat \rho^{(1)}_{0,ACA(n)}],
\end{align}
\end{subequations}
where we have chosen a single-site local approximation\mods{, i.e., $N_\mathrm{int}=2$} and $n$ determines the order of the ACA. The resulting $L$ functionals $F^{U\hat n_ {i,\uparrow} \hat n_{i,\downarrow}}[\mat \rho^{(1)}_{0,ACA(n)}]$ to be computed consider only $2(n+1)$ one-particle states instead of $2L$ of the full system. The results discussed in Sec.~\ref{sec:ACA} show that the first-order ACA, i.e. $n=1$ already gives results close to the exact value and quantum computers with at least 4 qubits are widely available.
Thus, this work will consider the evaluation of the RDMF $F^{\hat W_1}[\mat \rho^{(1)}_{0,ACA(1)}]$ with $\hat W_1=U \hat n_ {1,\uparrow} \hat n_{1,\downarrow}$ as an exemplary case.

\subsection{\label{sec:results_trial_state}Trial State}
\begin{figure}
    \centering
\scalebox{0.65}{
\Qcircuit @C=1.0em @R=0.2em @!R { \\
                \nghost{ {q}_{0} :  } & \lstick{ {q}_{0} :  } & \gate{\mathrm{R_Z}\,({u_0})} & \gate{\mathrm{\sqrt{X}}} & \gate{\mathrm{R_Z}\,({u_4})} & \ctrl{1} & \gate{\mathrm{R_Z}\,({u_{8}})} & \gate{\mathrm{\sqrt{X}}} & \gate{\mathrm{R_Z}\,({u_{12}})} & \qw & \qw & \qw\\ 
                \nghost{ {q}_{1} :  } & \lstick{ {q}_{1} :  } & \gate{\mathrm{R_Z}\,({u_1})} & \gate{\mathrm{\sqrt{X}}} & \gate{\mathrm{R_Z}\,({u_5})} & \targ & \ctrl{1} & \gate{\mathrm{R_Z}\,({u_9})} & \gate{\mathrm{\sqrt{X}}} & \gate{\mathrm{R_Z}\,({u_{13}})} & \qw & \qw\\ 
                \nghost{ {q}_{2} :  } & \lstick{ {q}_{2} :  } & \gate{\mathrm{R_Z}\,(\mathrm{u_2})} & \gate{\mathrm{\sqrt{X}}} & \gate{\mathrm{R_Z}\,(\mathrm{u_6})} & \ctrl{1} & \targ & \gate{\mathrm{R_Z}\,({u_{10}})} & \gate{\mathrm{\sqrt{X}}} & \gate{\mathrm{R_Z}\,({u_{14}})} & \qw & \qw\\ 
                \nghost{ {q}_{3} :  } & \lstick{ {q}_{3} :  } & \gate{\mathrm{R_Z}\,(\mathrm{u_3})} & \gate{\mathrm{\sqrt{X}}} & \gate{\mathrm{R_Z}\,({u_7})} & \targ & \gate{\mathrm{R_Z}\,({u_{11}})} & \gate{\mathrm{\sqrt{X}}} & \gate{\mathrm{R_Z}\,({u_{15}]})} & \qw & \qw & \qw\\ 
}}
    \caption{Hardware efficient trial state used in this work with 16 real parameters $\mat u$.}
    \label{fig:trial_state}
\end{figure}
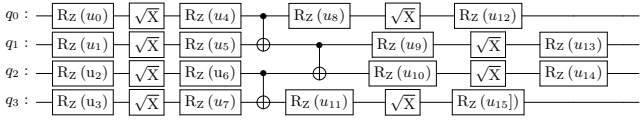

As the hardware-efficient trial state we propose the variant shown in Fig.~\ref{fig:trial_state} tailored to IBM quantum computers with the Falcon r4L processor type which have $R_z$-, $\sqrt{X}$- and CNOT-gates as basic gates and a linear topology for basic two-qubit gates~\cite{ibmq}.
The Bravyi-Kitaev transformation is used for the fermionic transformation and the two interacting one-particle states are placed on the two center qubits.

The chosen hardware efficient trial state is minimal in terms of the number of two-qubit gates. We were not able to find a parametrization of the trial state that represents the one-particle reduced density matrix $\mat \rho^{(1)}_{0,ACA(1)}$ of the chosen example case without the additional freedom generated by the additional unitary transform of the non-interacting states introduced in Sec.~\ref{sec:flex}. One exception is the case of infinite interaction strengths, i.e. $U/t\rightarrow \infty$, where the ACA-transformed density matrix $\mat \rho^{(1)}_{0,ACA(1)}$ is diagonal. Hence, this additional freedom makes the constraints in the constrained minimization problem satisfiable for arbitrary interactions strengths and allows us to use such a simple trial state. 
The implementation of the augmented Lagrangian and input files for the presented results are publicly available~\cite{robert_schade_2021_5749768}.

\subsection{\label{sec:results_no_noise}Simulation without Noise}
To show the suitability of the chosen trial state the constrained minimization with the augmented Lagrangian was run without noise, i.e. exact expectation values. Details for the constrained minimization can be found in appendix~\ref{sec:auglag_details}. Fig.~\ref{fig:conv} shows the convergence of the value of the augmented Lagrangian $L$ and the expectation value of the interaction $\langle \hat W_1\rangle=U \hat n_{1,\uparrow} \hat n_{i,\downarrow}$, i.e. the value of the RDMF in comparison to the value $F_\mathrm{exact}$ of the RDMF obtained from a constrained minimization over Slater determinants (see appendix~\ref{sec:exact_functional_details}). Additionally, the overall constraint violation ${\sum_i c_i^2}$ is shown. 
The convergence of the augmented Lagrangian is rapid so that typically no more than 10 outer iterations are required.
If the Lagrange multipliers are initialized to zero, i.e. $\mat \lambda_0=0$, then the augmented Lagrangian approach converges from a large constraint violation to the final result. However, if an initial guess for the Lagrange multipliers is obtained from the Müller functional~\cite{MULLER1984446}, then the constraint violation is already small after the first iteration of the augmented Lagrangian and the convergence of the interaction energy $\langle \hat W \rangle$ is much quicker.
The Müller functional~\cite{MULLER1984446} is an approximate parametrized RDMF and its derivatives with respect to the one-particle reduced density matrix, which correspond to the Lagrange multipliers, can be obtained with polynomial cost.
\begin{figure}
    \centering
    \includegraphics[width=0.5\textwidth]{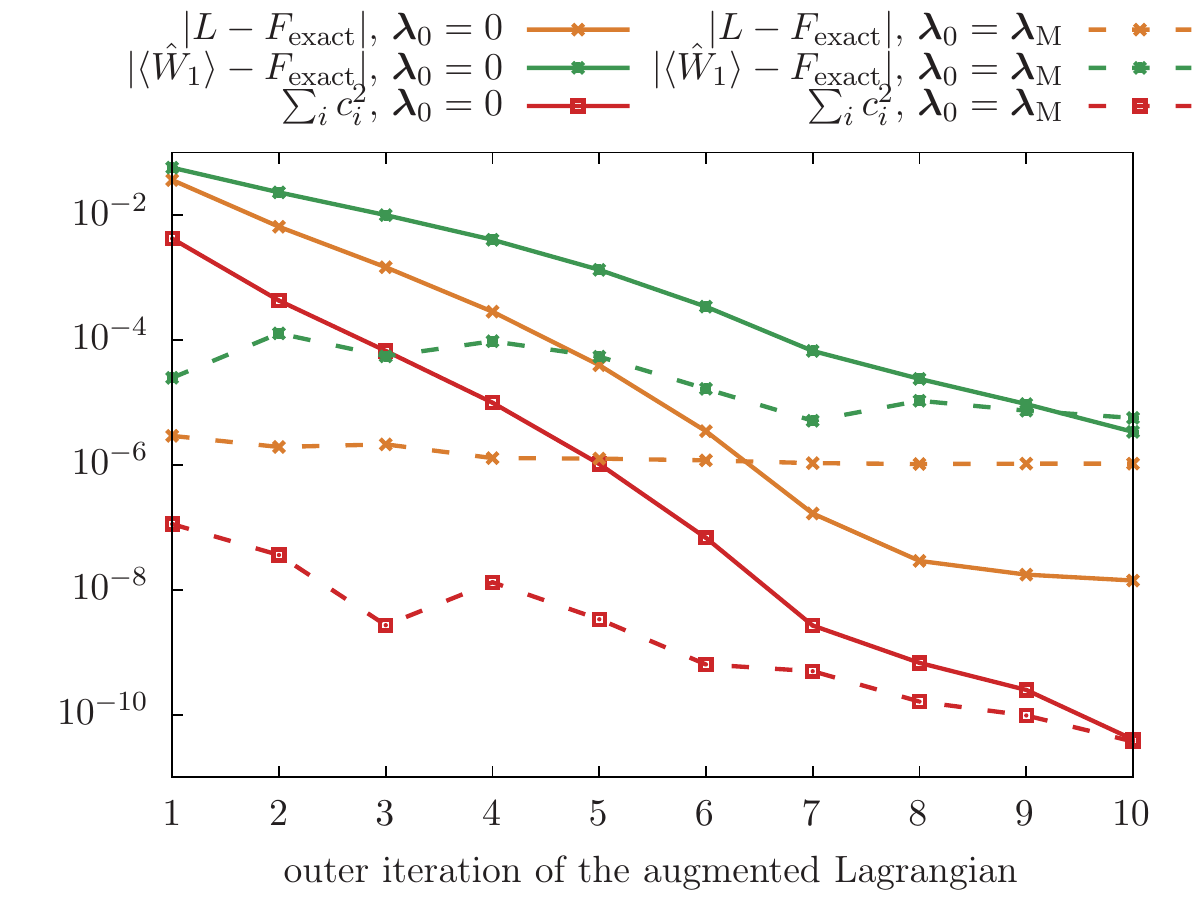}
    \caption{Convergence of the augmented Lagrangian $L$, the interaction energy $\langle\hat W_1\rangle=U \hat n_{1,\uparrow} \hat n_{i,\downarrow}$ and the constraint violation ${\sum_i c_i^2}$ at the end of each solution of an unconstrained subproblem \mods{in a noiseless simulation} for the situation defined in Sec.~\ref{sec:results_local_aca} and $U/t=1$. $F_\mathrm{exact}$ is the value of the RDMF obtained with a constrained minimization over Slater determinants (see appendix~\ref{sec:exact_functional_details}). Solid lines show the convergence when the initial Lagrange multipliers are chosen as zero and dashed lines the corresponding results if the initial values of the Lagrange multipliers are chosen as the derivatives of the Müller functional.}
    \label{fig:conv}
\end{figure}

\begin{figure}
    \centering
    \includegraphics[width=0.5\textwidth]{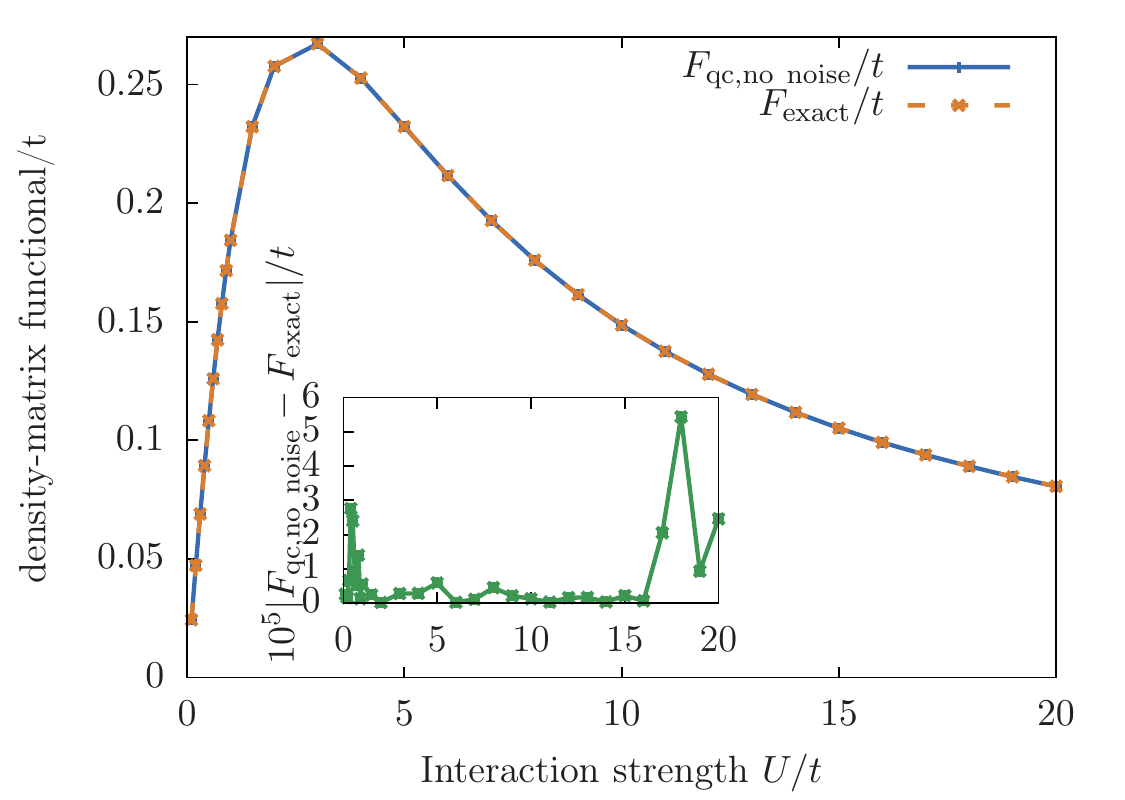}
    \caption{Comparison of the RDMF $F_\mathrm{qc,no\ noise}$ obtained with the augmented Lagrangian and the trial state in Fig.~\ref{fig:trial_state} without noise compared to the exact value $F_\mathrm{exact}$ (see appendix~\ref{sec:exact_functional_details}) for the RDMF $F^{U\hat n_ {1,\uparrow} \hat n_{1,\downarrow}}[\mat \rho^{(1)}_{0,ACA(1)}]$ defined in Sec.~\ref{sec:results_local_aca}.}
    \label{fig:U}
\end{figure}

The results for the RDMF $F_\mathrm{qc,no\ noise}$ obtained with the augmented Lagrangian and the trial state in Fig.~\ref{fig:trial_state} without noise are shown for  a wide range of interactions strengths $U/t$ in Fig.~\ref{fig:U}. The chosen trial state together with the unitary transformation of the non-interacting states introduced in Sec.~\ref{sec:flex} obtains the density-functional that is nearly indistinguishable from the exact result for arbitrary $U/t$, i.e. in the weakly interacting regime $U/t\ll1$, as well as in the strongly interacting and correlated regime $U/t\gg 1$.

\subsection{\label{sec:results_device}Results on NISQ Device}
After showing that the proposed schema with the chosen trial state converges accurately to the RDMF when exact expectation values are considered, we now include noise into the simulation. 

We focus on IBM quantum computers~\cite{ibmq} with a linear coupling topology like the ibmq\_bogota, ibmq\_manila, or ibmq\_santiago machines, respectively. These are all 5-qubit quantum computers with the processor type Falcon r4L.
The noisy simulations have been performed with the density-matrix-based simulator in Qiskit 0.29.0 using the noise model of the ibmq\_manila quantum computer, the qubit-wise commutation of measurements, and 8192 shots per measurement. The noise simulation included depolarizing gate errors, thermal relaxation errors, and single-qubit readout errors. The first four qubits of the 5-qubit quantum computer were used.

\begin{figure}
    \centering
    \includegraphics[width=0.5\textwidth]{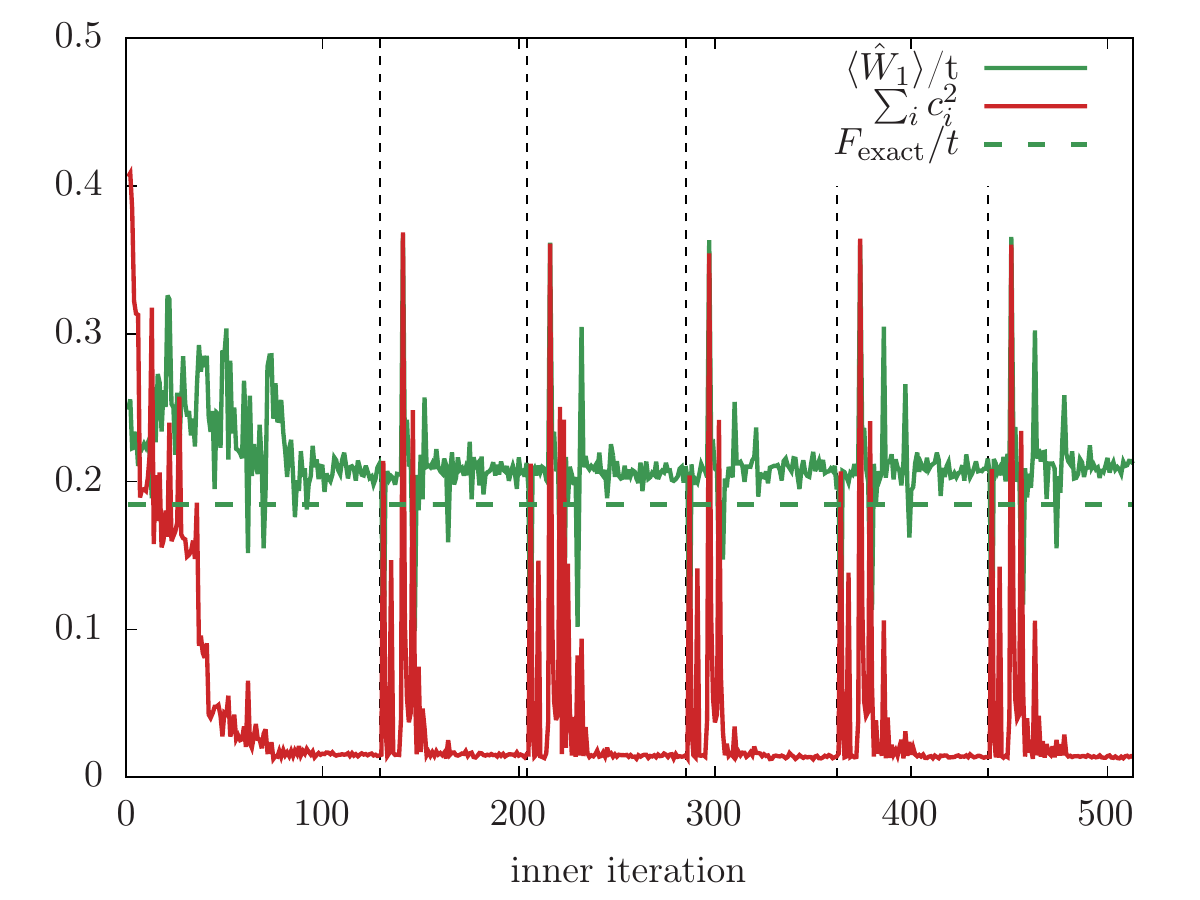}
    \caption{Convergence of the expectation value $\langle \hat W_1\rangle=U \hat n_{1,\uparrow} \hat n_{i,\downarrow}$ of the interaction and and the overall constraint violation $\sum_i c_i^2$ during the solution of the unconstrained subproblems of the augmented Lagrangian for the RDMF $F^{U\hat n_ {1,\uparrow} \hat n_{1,\downarrow}}[\mat \rho^{(1)}_{0,ACA(1)}]$ defined in Sec.~\ref{sec:results_local_aca} for $U/t=1$. The expectation values of the trial states were obtained from a noisy simulation on a classical computer. \mods{The corresponding results obtained on a  quantum computer are shown in Fig.~\ref{fig:conv_noise_real}}. Vertical dashed lines indicate different outer iterations, i.e., updates of the penalty parameters and Lagrange multipliers.}
    \label{fig:conv_noise_sim}
\end{figure}

Fig.~\ref{fig:conv_noise_sim} shows the behavior of the expectation value of the interaction $\langle \hat W_1\rangle$ and the constrained violation $\sum_i c_i^2$ during the constrained minimization when the quantum programs are simulated with the noise model of the real quantum computer ibmq\_manila. The corresponding calibration results are given in table~\ref{tab:ibmq_manila}.
The convergence when using the real quantum computer to evaluate the expectation values during the minimization is very similar to the noisy simulation. 

During the first outer iteration of the augmented Lagrangian, the constraint violation converges in both the noisy simulation and real execution to a finite value. For a solution of a constrained minimization problem with the augmented Lagrangian approach this behavior is not unexpected, as can be seen also in Fig.~\ref{fig:conv}. However, after the penalty update, i.e. increase of the penalties on the constraints, the convergence to essentially the same constraint violation in the second and subsequent outer iterations is curious and points to a deeper issue. Two possible explanations exist: Either the minimization in the first outer iteration has converged to a local minimum that is not the global minimum and subsequent iterations are not able to escape this minimum, or due to the inclusion of noise, the trial state can not represent a state that fulfills the constraints. The dominant part of the overall constraint violation stems from the constraints on the off-diagonal elements $\rho^{(1)}_{ACA(1),1\uparrow,2\uparrow}$ and $\rho^{(1)}_{ACA(1),1\downarrow,2\downarrow}$, respectively, i.e. the covalencies.
\begin{figure}
    \centering
    \includegraphics[width=0.5\textwidth]{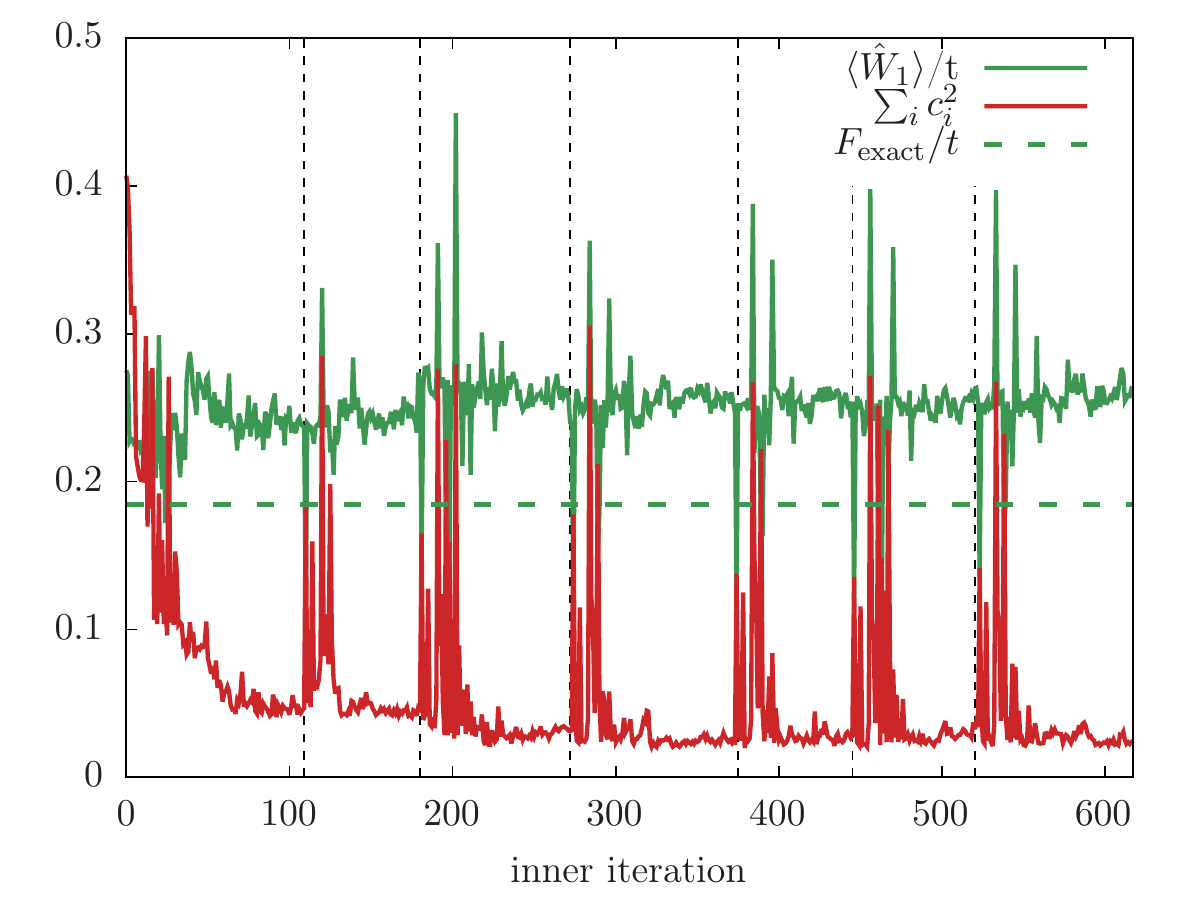}
    \caption{Convergence of the expectation value $\langle \hat W_1\rangle=U \hat n_{1,\uparrow} \hat n_{i,\downarrow}$ of the interaction and the overall constraint violation $\sum_i c_i^2$ during the solution of the unconstrained subproblems of the augmented Lagrangian for the RDMF $F^{U\hat n_ {1,\uparrow} \hat n_{1,\downarrow}}[\mat \rho^{(1)}_{0,ACA(1)}]$ defined in Sec.~\ref{sec:results_local_aca} for $U/t=1$ when executed with the ibmq\_manila quantum computer. Vertical dashed lines indicate different outer iterations, i.e. updates of the penalty parameters and Lagrange multipliers.}
    \label{fig:conv_noise_real}
\end{figure}

\begin{figure}
    \centering
    \includegraphics[width=0.5\textwidth]{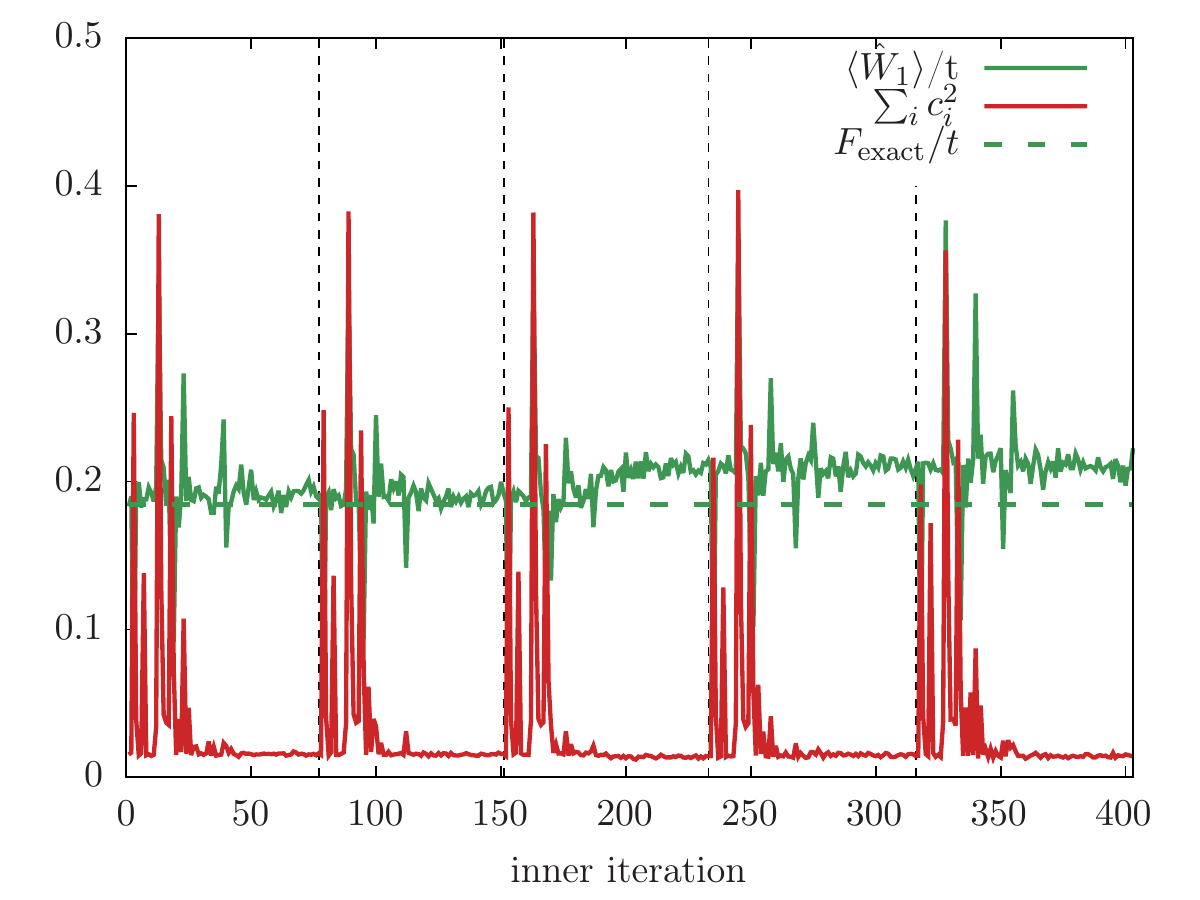}
    \caption{Convergence for the same situation as shown in Fig.~\ref{fig:conv_noise_sim} but starting from the converged parameters $\mat u_\mathrm{noiseless}$ of the noiseless simulation.}
    \label{fig:conv_noise_sim_from_noiseless}
\end{figure}

Using the converged parameters $\mat u_\mathrm{noiseless}$ of the trial state from the noiseless constrained minimization as a starting point for the constrained minimization with noise, the convergence as shown in Fig.~\ref{fig:conv_noise_sim_from_noiseless} is obtained. The converged values are very similar to the results shown in Fig.~\ref{fig:conv_noise_sim}, where the minimization commenced from a random starting point. Thus, we conclude that the convergence to a local minimum rather than a global minimum is very unlikely and that the observed behavior points to a representability issue of the many-particle state on the noisy quantum computer.
This aspect is the subject of future research.

We therefore conclude that the evaluation of the RDMF is possible on a quantum computer. The results obtained this way can then be used in an iterative total energy minimization. The formulation of the total energy minimization problem with the RDMF allows us to employ novel approximations like the ACA to drastically reduce the qubit requirements of the problem and to parallelize it.

\section{\label{sec:conclusion}Conclusions and Outlook}
We have introduced a hybrid quantum-classical approach based on an RDMFT formulation of the quantum total energy problem and the ACA for gate-based quantum computers.
Using the latter, a drastic reduction of the necessary qubit count is demonstrated.
The measurements of the one-particle reduced density matrix that are required for the density-matrix constraints are shown to be obtainable with only a linear number of quantum programs when the general commutativity of observables is exploited. A construction algorithm for the measurement programs is given and available degrees of freedom for their optimization are introduced.

The \mods{essential part of the proposed approach, the evaluation of a local RDMF,} is demonstrated with a Hubbard-type model system using noise-free simulations, simulations including noise models of realistic quantum computers, as well as with executions on real IBM quantum computers. While model systems such as the Hubbard model are convenient to study the features and convergence behavior of the proposed algorithm, the goal of this approach is the application within \textit{ab-initio} molecular dynamics calculations. The main difference between the treatment of model systems and realistic systems is the number of terms in the interaction Hamiltonian. The variational formulation based on RDMFT makes the proposed algorithm very suitable for \textit{ab-initio} molecular dynamics calculations because the forces can be evaluated in a straightforward way from already available quantities such as the one- and two-particle reduced density matrix~\cite{PhysRevLett.98.066401}. Noise in the nuclear forces stemming from the noise of the quantum computer can be compensated for in \textit{ab-initio} molecular dynamics simulations in the spirit of approximate computing~\cite{computation8020039}, where the desired thermodynamic expectation values can nevertheless be accurately obtained by devising a properly modified Langevin equation~\cite{kuhne2018disordered,PhysRevResearch.3.013125}. The investigation of the representability of fermionic quantum states on noisy gate-based quantum computers, the optimization of the measurement programs as well as the integration with molecular dynamics programs like CP2K~\cite{doi:10.1063/5.0007045} are subject for future research.

\section{\label{sec:acknowledgements}Acknowledgements}
The authors acknowledge the use of IBM Quantum services for this work. The views expressed are those of the authors, and do not reflect the official policy or position of IBM or the IBM Quantum team. In this paper the ibmq\_manila which is one of the IBM Quantum Falcon Processors has been used. The authors gratefully acknowledge the funding of this project by computing time provided by the Paderborn Center for Parallel Computing (PC²). This work is partially funded by Paderborn University’s research award for ``GreenIT'', as well as the Federal Ministry of Education and Research (BMBF) and the state of North Rhine-Westphalia as part of the NHR Program.

\appendix

\section{Availability of Source Code and Data}\label{sec:source}
The source code of the software implementation of the presented approach, input files for the presented results as well as the raw data are openly available at~\cite{robert_schade_2021_5749768}.

\section{Practical Details of the Exact RDMF}\label{sec:exact_functional_details}
The exact reference of the RDMF has been obtained by parametrizing 
\begin{equation}
    F_\mathrm{exact}^{\hat W}[\mat \rho^{(1)}]= \min_{\mat x \in \mathbb{C}^{2^{N_\chi}}} \langle \Psi (\mat x)|\hat W|\Psi(\mat x)\rangle
\end{equation}
with the equality constraints
\begin{subequations}
\begin{align}
    \langle \Psi(\mat x)|\hat c^\dagger_\alpha \hat c_\beta|\Psi(\mat x)\rangle&=\rho^{(1)}_{\beta,\alpha},\\
    \langle \Psi(\mat x)|\Psi(\mat x)\rangle&=1,
\end{align}
\end{subequations}
where $|\Psi(\mat x)\rangle=\sum_i x_i |n_i\rangle$ is a sum over all Slater determinants $|00 \cdots\rangle$, $|10\cdots\rangle$, ..., respectively.
The complex parameters $x_i$ were represented by their real and imaginary parts and the constrained minimization was performed using the trust-region constrained minimization algorithm from SciPy 1.7.1~\cite{scipy,doi:10.1137/1.9780898719857} with a convergence tolerance of $10^{-9}$. 

\section{Practical Details of the Augmented Lagrangian Approach}\label{sec:auglag_details}
\paragraph{Parameters for the Augmented Lagrangian}
The initial values for the parameters $\mat u_0$ were uniformly chosen at random from $[-\pi,\pi]$. The initial values for the Lagrange multipliers $\mat \lambda_0$ were either chosen as zero, or as the numerical derivatives of the Müller functional~\cite{MULLER1984446} for the given one-particle reduced density matrix. The initial value for the penalties $\mat\mu_0$ was chosen as $10$ and the penalties were updated after each solution of the unconstrained problem by multiplying them with $1.5$.

\subsection{Solution of the Unconstrained Subproblems}
In the case without noise, the unconstrained subproblems were solved with the L\_BFGS\_B and COBYLA algorithms implemented in Qiskit 0.29.0~\cite{Qiskit,scipy}. The tolerance was set to $10^{-3}$ and at most $10000$ iterations were permitted. The optimization of the unitary transform of the non-interacting one-particle states outlined in Sec.~\ref{sec:flex} has been performed with the BFGS algorithm (with a convergence tolerance of $10^{-9}$) from SciPy 1.7.1~\cite{scipy} and the parametrization of unitary matrices
\begin{equation}
    U=e^{iH},
\end{equation}
where $H$ is a parametrized arbitrary hermitian matrix.

In the case with noise, the COBYLA algorithm implemented in Qiskit 0.29.0~\cite{Qiskit,scipy} was used with a convergence tolerance of $0.01$.

\section{Calibration of the ibmq\_manila Quantum Computer}\label{sec:appendix_qc}
The thermal relaxation time constant T$_1$, the dephasing time constant $T_2$, and error rates from calibration of the ibmq\_manila quantum computer are shown in table~\ref{tab:ibmq_manila}.
\onecolumngrid

\begin{table}[]
    \centering
\begin{widetext}
\begin{tabular}{c|c|c|c|c|c}
    Qubit & Q0 & Q1 & Q2 & Q3 & Q4 \\ \hline
    T$_1$ in $\mu$s & 128.51 &142.79 & 184.95 & 214.35 & 127.79 \\
    T$_2$ in $\mu$s & 74.34 & 79.95 & 24.85 & 69.18 & 44.4 \\
    Frequency in GHz & 4.963 & 4.838 & 5.037 & 4.951 & 5.066 \\
    Anharmonicity in GHz & -0.34335 & -0.34621 & -0.34366 & -0.34355 & -0.34211 \\
    Readout error & 0.0259 & 0.0324 & 0.0190 & 0.02210 & 0.01920 \\
    SPAM error $|1\rangle \rightarrow |0\rangle$ & 0.0438 & 0.0482 & 0.0286 & 0.0306 & 0.0302 \\
    SPAM error $|0\rangle \rightarrow |1\rangle$ & 0.0080 & 0.0166 & 0.0094 & 0.0136 & 0.0082 \\
    Readout length in ns & 5351.111 & 5351.111& 5351.111& 5351.111& 5351.111 \\
    Identity,$\sqrt{x}$,$x$ errors $\cdot 10^4$ &  1.920 & 3.970 & 2.571 & 1.834 & 3.155 \\
    CNOT error$\cdot 10^3$ & 0 to 1 6.684 & 1 to 2 9.947 & 2 to 3 7.344 & 3 to 4 4.746 & \\
    CNOT Gate time in ns & 277.33 & 312.889 & 504.889 & 391.111 & 298.667 \\
\end{tabular}
\end{widetext}
    \caption{Calibration used for the ibmq\_manila quantum computer.}
    \label{tab:ibmq_manila}
\end{table}
\twocolumngrid

\bibliography{lit}
\end{document}